\documentstyle[]{article}
\catcode`\@=11 

\def\@aabuffer{}
\def\author #1{\expandafter\def\expandafter\@aabuffer\expandafter
{\@aabuffer \small\rm      #1\relax \par}}
\def\address#1{\expandafter\def\expandafter\@aabuffer\expandafter
{\@aabuffer \small\it #1\relax \par\vspace{1em}}}

\def\maketitle{
\begin{center}
   {\bf \@title \par}       

   \vskip 2em                      
   \@aabuffer\relax
\end{center} \par
\gdef\@aabuffer{}
}

\def\abstracts#1{
\begin{center}
{\begin{minipage}{4.2truein}
                 \footnotesize
                 \parindent=0pt #1\par
                 \end{minipage}}\end{center}
                 \vskip 2em \par}

\fussy
\def\section{\@startsection {section}{1}{\z@}{-3.5ex plus -1ex minus 
    -.2ex}{2.3ex plus .2ex}{\bf }}
\def\subsection{\@startsection{subsection}{2}{\z@}{-3.25ex plus -1ex  
minus 
   -.2ex}{1.5ex plus .2ex}{\it }}

\def\@makefnmark{{$\!^{\@thefnmark}$}}

\renewenvironment{thebibliography}[1]
	{\begin{list}{\arabic{enumi}.}
	{\usecounter{enumi}\setlength{\parsep}{0pt}
	 \setlength{\itemsep}{0pt} 
         \settowidth
	{\labelwidth}{#1.}\sloppy}}{\end{list}}

\newcounter{arabiclistc}

\def\@citex[#1]#2{\if@filesw\immediate\write\@auxout
	{\string\citation{#2}}\fi
\def\@citea{}\@cite{\@for\@citeb:=#2\do
	{\@citea\def\@citea{,}\@ifundefined
	{b@\@citeb}{{\bf ?}\@warning
	{Citation `\@citeb' on page \thepage \space undefined}}
	{\csname b@\@citeb\endcsname}}}{#1}}

\newif\if@cghi
\def\cite{\@cghitrue\@ifnextchar [{\@tempswatrue
	\@citex}{\@tempswafalse\@citex[]}}
\def\citelow{\@cghifalse\@ifnextchar [{\@tempswatrue
	\@citex}{\@tempswafalse\@citex[]}}
\def\@cite#1#2{{$\!^{#1}$\if@tempswa\typeout
	{IJCGA warning: optional citation argument 
	ignored: `#2'} \fi}}

\def\baselinestretch{1.0}
\ifx\selectfont\undefined
\@normalsize\else\let\glb@currsize=\relax\selectfont
\fi

\ifx\selectfont\undefined
\def\@singlespacing{%
\def\baselinestretch{1}\ifx\@currsize\normalsize\@normalsize\else\@cur 
rsize\fi%
}
\else
\def\@singlespacing{\def\baselinestretch{1}\let\glb@currsize=\relax\se 
lectfont}
\fi

\long\def\@makecaption#1#2{
   \vskip 10pt 
   \setbox\@tempboxa\hbox{\footnotesize #1: #2}
   \ifdim \wd\@tempboxa >\hsize   
       \leftskip 0pt plus 1fil 
       \rightskip 0pt plus -1fil 
       \parfillskip 0pt plus 2fil 
       \footnotesize #1: #2\par   
     \else                        
       \hbox to\hsize{\hfil\box\@tempboxa\hfil}  
   \fi}

\def\Journal#1#2#3#4{{#1} {\bf #2}, #3 (#4)}
\def\Book#1#2#3{{\bf #1}, {#2} (#3)}

\def\NPB{{\em Nucl. Phys.} B}
\def\PLB{{\em Phys. Lett.} B}
\def\PRL{\em Phys. Rev. Lett.}

\def\PR{{\em Phys. Rev.}}

\def\PHYSICA{{\em Physica} D}
\def\CMP{\em Commun. Math. Phys.}
\def\PREP{\em Phys. Rep.}
\def\JMP{\em J. Math. Phys.}
\def\CQG{\em Class. Quant. Grav.}
\def\ANN{\em Annals of Physics}

\def\mco{\multicolumn}

\def\be{\begin{equation}}
\def\ee{\end{equation}}
\def\bea{\begin{eqnarray}}
\def\eea{\end{eqnarray}}
\def\bd{\begin{displaymath}}
\def\ed{\end{displaymath}}
\def\mymatrix#1{\null\,\vcenter{\normalbaselines\m@th
  \ialign{\hfil$##$&&\quad\hfil$##$\crcr
	\mathstrut\crcr\noalign{\kern-\baselineskip}
	#1\crcr\mathstrut\crcr\noalign{\kern-\baselineskip}}}\,}
\catcode`\@=12 
\global\newcount\fno \global\fno=0

\def\bra#1{\left\langle #1\right|}
\def\ket#1{\left| #1\right\rangle}

\def\p{{\scriptscriptstyle +}}
\def\m{{\scriptscriptstyle -}}
\def\grad#1{\,\nabla\!_{{#1}}\,}

\def\lform{\hbox{$\sqcup$}\llap{\hbox{$\sqcap$}}}
\def\darr#1{\raise1.5ex\hbox{$\leftrightarrow$}\mkern-16.5mu #1}

\def\half{{{\scriptstyle 1}\over{\scriptstyle 2}}} 
\def\threehalf{{{\scriptstyle 3}\over{\scriptstyle 2}}} 
\def\Ntwo{{{\scriptstyle N}\over{\scriptstyle 2}}} 
\def\quarter{{{\scriptstyle 1}\over{\scriptstyle 4}}} 
\def\third{{{\scriptstyle 1}\over{\scriptstyle 3}}} 
\def\roughly#1{\raise.3ex\hbox{$#1$\kern-.75em\lower1ex\hbox{$\sim$}}}

\def\e#1{{\rm e}^{#1}}
\def\p{\raise2pt\hbox to5pt{\fiverm +}}
\def\m{\raise2pt\hbox to5pt{\fiverm --}}
\def\grad#1{\,\nabla\!_{{#1}}\,}

\def\lform{\hbox{$\sqcup$}\llap{\hbox{$\sqcap$}}}
\def\darr#1{\raise1.5ex\hbox{$\leftrightarrow$}\mkern-16.5mu #1}

\def\roughly#1{\raise.3ex\hbox{$#1$\kern-.75em\lower1ex\hbox{$\sim$}}}
\def\BBox{\Box}
\def\littlebox{\lform}
\def\tr{{\rm tr}\, }
\def\barJ{\bar{J}}
\def\barpsi{\bar{\psi}}
\def\barsigma{\bar{\sigma}}
\def\barchi{\bar{\chi}}
\def\barlam{\bar{\lambda}}
\def\bartheta{\bar{\theta}}
\def\barxi{\bar{\xi}}

\def\barPhi{\overline{\Phi}}
\def\barQ{\overline{Q}}
\def\barW{\overline{W}}
\def\barphi{\bar{\phi}}
\def\metric{\eta^{mn}}

\def\alphadot{\dot{\alpha}}
\def\betadot{\dot{\beta}}
\def\gammadot{\dot{\gamma}}
\def\deltadot{\dot{\delta}}
\def\onedot{\dot{1}}
\def\twodot{\dot{2}}
\def\psiu{\psi^{\alpha}}
\def\psiudot{\barpsi^{\alphadot}}
\def\psid{\psi_{\alpha}}
\def\psiddot{\barpsi_{\alphadot}}
\def\epsu{\epsilon^{\alpha\beta}}
\def\epsudot{\epsilon^{\alphadot\betadot}}
\def\epsd{\epsilon_{\alpha\beta}}
\def\epsddot{\epsilon_{\alphadot\betadot}}
\def\sig{\sigma^{m}_{\alpha\betadot}}
\def\sigdot{\barsigma^{m\alphadot\beta}}
\def\barD{\overline{D}}
\def\D{D_{\alpha}}
\def\Ddot{\barD_{\alphadot}}
\def\Dt{\tilde{D}}
\def\barDt{\overline{\Dt}}
\def\Dta{\Dt _{\alpha}}
\def\Ddott{\barDt _{\alphadot}}
\def\W{W_{\alpha}}
\def\Wdot{\barW_{\alphadot}}
\def\ni{\noindent}
\def\poin{Poincar\'e}
\def\cliff{\ket{\Omega}}
\def\th{\theta}
\def\barth{\overline{\theta}}
\def\tha{\theta ^{\alpha}}
\def\bartha{\overline{\theta}_{\alphadot}}
\def\tht{\tilde{\theta}}
\def\bartht{\overline{\tilde{\theta}}}
\def\that{\tht ^{\alpha}}
\def\barthat{\bartht_{\alphadot}}
\def\dd{\partial _{\alpha}}
\def\du{\partial ^{\alpha}}
\def\bard{\overline{\partial}}
\def\bardd{\overline{\partial}_{\alphadot}}
\def\bardu{\overline{\partial}^{\alphadot}}
\def\sx{(x,\th ,\barth )}
\def\sy{(y,\th )}
\def\sxt{(x,\th ,\barth ,\tht ,\bartht )}
\def\yt{\tilde{y}}
\def\syt{(\yt ,\th )}
\def\dx{\delta _{\xi}}
\def\dff{\int d^4x\!\!\int d^4\th\;}
\def\dft{\int d^4x\!\!\int d^2\th\;}
\def\df{\int d^4x\,}
\def\dtf{\int d^4\th\;}
\def\dt{\int d^2\th\;}

\def\dtt{\int d^2\th d^2\tht\;}
\def\lag{{\cal L}}
\def\Kahler{K\"ahler}
\def\leaderfill{\leaders\hbox to 1em{\hss.\hss}\hfill}
\def\tr{{\rm tr}\,}
\def\tym{\theta _{\rm YM}}
\def\pp{{\cal F}}
\def\ppf{{\cal F}_{\rm eff}}

\begin{document}
\hbox{\ }
\vskip -1.2in\vbox{\kern -1in\hbox{\hfil\hskip 3.75in FERMILAB-PUB-96/445-T}}
\vskip 1in

\title{INTRODUCTION TO SUPERSYMMETRY}

\author{ JOSEPH D. LYKKEN }

\address{Fermi National Accelerator Laboratory\\
P.O. Box 500\\
Batavia, IL 60510}


\maketitle\abstracts{
These lectures give a self-contained introduction to supersymmetry from a
modern perspective. Emphasis is placed on material essential to understanding
duality. Topics include: central charges and BPS-saturated states,
supersymmetric nonlinear sigma models, N=2 Yang-Mills theory, holomorphy and
the N=2 Yang-Mills $\beta$ function, supersymmetry in 2, 6, 10, and 11
spacetime dimensions.
}

\section{Introduction}

\begin{quote}
{\it ``Never mind, lads. Same time tomorrow. We must get a winner
one day.''}\\ -- Peter Cook, as the doomsday prophet in
``The End of the World''.
\end{quote}

Supersymmetry, along with its monozygotic sibling superstring theory,
has become the dominant framework for formulating physics beyond
the standard model. This despite the fact that, as of this morning,
there is no unambiguous experimental evidence for either idea.
Theorists find supersymmetry appealing for reasons which are both
phenomenological and technical. In these lectures I will focus
exclusively on the technical appeal. There are many good recent
reviews of the phenomenology of supersymmetry. \cite{report}
Some good technical reviews are
Wess and Bagger, \cite{wb} West, \cite{west} and Sohnius. \cite{soh}

The goal of these lectures is to provide the student with
the technical background requisite for the recent applications
of duality ideas to supersymmetric gauge theories and superstrings.
More specifically, if you absorb the material in these lectures,
you will understand Section 2 of Seiberg and Witten, \cite{wit}
and you will have a vague notion of why there might be such a
thing as $M$-theory. Beyond that, you're on your own.

\section{Representations of Supersymmetry}

\subsection{The general 4-dimensional supersymmetry algebra}
\label{subsec:galgebra}

A symmetry of the S-matrix means that the symmetry transformations
have the effect of merely reshuffling the asymptotic single
and multiparticle states. The known symmetries of the S-matrix
in particle physics are:
\begin{itemize}
\item \poin\ invariance, the semi-direct product of
translations and Lorentz rotations, with generators
$P_m$, $M_{mn}$.
\item So-called ``internal'' global symmetries, related to
conserved quantum numbers such as electric charge and
isospin. The symmetry generators are Lorentz scalars and
generate a Lie algebra,
\be
\left[ B_{\ell}, B_{k} \right] = i C_{\ell k}^j B_j \quad ,
\label{eq:basiclie}
\ee
where the $C_{\ell k}^j$ are structure constants.
\item Discrete symmetries: C, P, and T.
\end{itemize}

In 1967, Coleman and Mandula \cite{coleman}
provided a rigorous argument
which proves that, given certain assumptions, the above
are the only possible symmetries of the S-matrix.
The reader is encouraged to study this classic paper
and think about the physical and technical assumptions which
are made there.

The Coleman-Mandula theorem can be evaded by weakening
one or more of its assumptions. In particular, the theorem
assumes that the symmetry algebra of the S-matrix involves
only commutators. Weakening this assumption to allow
anticommuting generators as well as commuting generators leads
to the possibility of {\bf supersymmetry}.
Supersymmetry (or SUSY for short) is defined as the
introduction of anticommuting symmetry generators
which transform in the $(\half,0)$ and $(0,\half)$
(i.e. spinor) representations of the Lorentz group.
Since these new symmetry generators are spinors, not scalars,
supersymmetry is not an internal symmetry. It is rather
an extension of the \poin\ spacetime symmetries.
Supersymmetry, defined as the extension of the \poin\
symmetry algebra by anticommuting spinor generators,
has an obvious extension to spacetime dimensions other
than four; the Coleman-Mandula theorem, on the other hand,
has no obvious extension beyond four dimensions.

In 1975, Haag, {\L}opusza\'nski, and Sohnius \cite{hls}
proved that supersymmetry is the only
additional symmetry of the S-matrix allowed by this
weaker set of assumptions. Of course, one could imagine
that a further weakening of assumptions might lead to
more new symmetries, but to date no physically compelling
examples have been exhibited. \cite{vandam}
This is the basis of the strong
but not unreasonable assertion that:
\vskip .2in{\bf \begin{center}
Supersymmetry is the only possible extension
of the known spacetime symmetries of particle physics.
\end{center}}\vskip .2in

In four-dimensional Weyl spinor notation (see the Appendix)
the N supersymmetry generators are denoted by
$Q_{\alpha}^A$, $A$$=$$1,\ldots$N. The most general
four-dimensional supersymmetry algebra is given in the Appendix;
here we will be content with checking some of the features of
this algebra.

The anticommutator of the $Q_{\alpha}^A$ with their adjoints
is:
\be
\{ Q^A_{\alpha},\barQ _{\betadot B}\} \quad =\quad
2\sigma ^{m}_{\alpha\betadot}P_{m}\delta ^A_B \quad .
\label{eq:qqbarcom}
\ee
To see this, note the right-hand side of Eq.~\ref{eq:qqbarcom}
must transform as $(\half,\half)$ under the Lorentz group.
The most general such object that can be constructed out of
$P_m$, $M_{mn}$, and $B_{\ell}$ has the form:
\bd
\sigma ^{m}_{\alpha\betadot}P_{m}C^A_B \quad ,
\ed
where the $C^A_B$ are complex Lorentz scalar coefficients.
Taking the adjoint of the left-hand side of Eq.~\ref{eq:qqbarcom},
using
\bea
\left( \sigma ^{m}_{\alpha\betadot} \right)^{\dagger} &=&
\sigma ^{m}_{\beta\alphadot} \quad ,\nonumber\\
\left( Q^A_{\alpha} \right)^{\dagger} &=&
\barQ ^A_{\alphadot} \quad ,
\label{eq:adjforms}
\eea
tells us that $C^A_B$ is a hermitian matrix. Furthermore,
since $\{ Q,\barQ \}$ is a positive definite operator,
$C^A_B$ is a positive definite hermitian matrix. This means
that we can always {\bf choose} a basis for the $Q^A_{\alpha}$
such that $C^A_B$ is proportional to $\delta ^A_B$. The
factor of two in Eq.~\ref{eq:qqbarcom} is simply a convention.

The SUSY generators $Q^A_{\alpha}$ commute with the translation
generators:
\be
[Q^A_{\alpha},P_m] = [\barQ ^A_{\alphadot},P_m]
 = 0 \quad .
\label{eq:transcom}
\ee
This is not obvious since the most general form
consistent with Lorentz invariance is:
\bea
\left[ Q^A_{\alpha},P_m\right] &=& Z^A_B \sigma _{\alpha\betadot m}
\barQ ^{\betadot B} \nonumber\\
{[} \barQ ^{\alphadot A},P_m ]
&=&\left(Z^A_B\right)^* Q^B_{\beta}
\barsigma _m^{\alphadot\beta} \quad ,
\label{eq:mostgtranscom}
\eea
where the $Z^A_B$ are complex Lorentz scalar coefficients.
Note we have invoked here the Haag, {\L}opusza\'nski, Sohnius
theorem which tells us that there are no $(\half,1)$ or
$(1,\half)$ symmetry generators.

To see that the $Z^A_B$ all vanish, the first step is
to plug Eq.~\ref{eq:mostgtranscom} into the Jacobi
identity:
\be
\left[ [Q^A_{\alpha},P_m],P_n \right] + ({\rm cyclic}) = 0
\quad .
\label{eq:jacid}
\ee
Using Eq.~\ref{eq:sltwocdef} this yields:
\be
-4i\left(ZZ^*\right)^A_B \sigma _{mn\alpha}^{\beta}
Q^B_{\beta} = 0 \quad ,
\label{eq:redjacid}
\ee
which implies that the matrix $ZZ^*$ vanishes.

This is not enough to conclude that $Z^A_B$ itself
vanishes, but we can get more information by considering
the most general form of the anticommutator of two $Q$'s:
\be
\{ Q_{\alpha}^A,Q_{\beta}^B \} = \epsd X^{AB} +
\epsilon_{\beta\gamma}\sigma _{\alpha}^{mn\gamma} M_{mn}
Y^{AB} \quad .
\label{eq:mostgenqq}
\ee
Here we have used the fact that the rhs must transform
as $(0,0) + (1,0)$ under the Lorentz group. The spinor
structure of the two terms on the rhs is
antisymmetric/symmetric respectively under
$\alpha\leftrightarrow\beta$, so the complex Lorentz scalar
matrices $X^{AB}$
and $Y^{AB}$ are also antisymmetric/symmetric respectively.

Now we consider $\epsu$ contracted on the Jacobi identity
\be
\left[ \{ Q_{\alpha}^A,Q_{\beta}^B \}, P_m \right]
+ \left\{ [P_m,Q_{\alpha}^A],Q_{\beta}^B \right\}
- \left\{ [Q_{\beta}^B,P_m],Q_{\alpha}^A \right\} = 0 \quad .
\label{eq:asjacobi}
\ee
Since $X^{AB}$ commutes with $P_m$, and plugging in
Eqs.~\ref{eq:qqbarcom},\ref{eq:mostgtranscom},\ref{eq:completeness}
and \ref{eq:raisecovsigma}, the above reduces to
\be
-4\left( Z^{AB} - Z^{BA} \right) P_m = 0 \quad ,
\label{eq:redjacobi}
\ee
and thus $Z^A_B$ is symmetric. Combined with $ZZ^*$$=$$0$
this means that $ZZ^{\dagger}$$=$$0$, which implies
that $Z^A_B$ vanishes, giving Eq.~\ref{eq:transcom}.

Having established Eq.~\ref{eq:transcom}, the symmetric part
of the Jacobi identity Eq.~\ref{eq:asjacobi} implies that
$M_{mn}Y^{AB}$ commutes with $P_m$, which can only be true if
$Y^{AB}$ vanishes. Thus:
\be
\{ Q_{\alpha}^A,Q_{\beta}^B \} = \epsd X^{AB} \quad .
\label{eq:betterqq}
\ee

The complex Lorentz scalars $X^{AB}$ are called {\bf central charges};
further manipulations with the Jacobi identities show that
the $X^{AB}$ commute with the $Q^A_{\alpha}$, $\barQ _{\alphadot A}$,
and in fact generate an Abelian invariant subalgebra of the
compact Lie algebra generated by $B_{\ell}$. Thus we can write:
\be
X^{AB} = a^{\ell AB}\; B_{\ell} \quad ,
\label{eq:breakccs}
\ee
where the complex coefficients $a^{\ell AB}$ obey the intertwining
relation Eq.~\ref{eq:intertwinerel}.

\subsection{The 4-dimensional N=1 supersymmetry algebra}

The Appendix also contains the special case of the
four-dimensional N=1 supersymmetry algebra. For N=1
the central charges $X^{AB}$ vanish by antisymmetry,
and the coefficients $S_{\ell}$ are real.
The Jacobi identity for $[[Q,B],B]$ implies that the
structure constants $C_{m\ell}^k$ vanish, so the internal
symmetry algebra is Abelian. Starting with
\bea
&[Q_{\alpha},B_{\ell}]  &=
S_{\ell} Q_{\alpha} \nonumber\\
&[\barQ _{\alphadot}, B_{\ell}]  &=
-S_{\ell} \barQ _{\alphadot} \quad ,
\label{eq:oldqbcom}
\eea
it is clear that we can rescale the Abelian generators $B_{\ell}$
and write:
\bea
&[Q_{\alpha},B_{\ell}]  &=
Q_{\alpha} \nonumber\\
&[\barQ _{\alphadot}, B_{\ell}]  &=
-\barQ _{\alphadot} \quad .
\label{eq:newqbcom}
\eea
Clearly only one independent combination of the Abelian
generators actually has a nonzero commutator with
$Q_{\alpha}$ and $\barQ _{\alphadot}$; let us denote this
$U(1)$ generator by $R$:
\bea
&[Q_{\alpha},R]  &=
Q_{\alpha} \nonumber\\
&[\barQ _{\alphadot}, R]  &=
-\barQ _{\alphadot} \quad .
\label{eq:rsymmcom}
\eea
Thus the N=1 SUSY algebra in general possesses an
internal (global) $U(1)$ symmetry known as {\bf R symmetry}.
Note that the SUSY generators have R-charge +1 and -1,
respectively.

\subsection{SUSY Casimirs}

Since we wish to characterize the irreducible representations
of supersymmetry on asymptotic single particle states, we
need to exhibit the Casimir operators. It suffices to do this
for the N=1 SUSY algebra, as the extension to N$>$1 is
straightforward.

Recall that the \poin\ algebra has two Casimirs: the
mass operator $P^2 = P_mP^m$, with eigenvalues $m^2$,
and the square of the Pauli-Ljubansk\'i vector
\be
W_m = \half \epsilon _{mnpq}P^nM^{pq} \quad .
\label{eq:pauliljub}
\ee
$W^2$ has eigenvalues $-m^2s(s+1)$, $s$$=$$0,\half ,1,\ldots$
for massive states, and $W_m = \lambda P_m$ for massless states,
where $\lambda$ is the helicity.

For N=1 SUSY, $P^2$ is still a Casimir (since $P$ commutes with
$Q$ and $\barQ$), but $W^2$ is not ($M$ does not commute with $Q$
and $\barQ$). The actual Casimirs are $P^2$ and $C^2$, where
\bea
C^2 &=& C_{mn}C^{mn} \quad ,\nonumber\\
C_{mn} &=& B_mP_n - B_nP_m  \quad , \\
B_m &=& W_m - \quarter \barQ _{\alphadot}
\barsigma _m^{\alphadot\beta}Q_{\beta} \quad .\nonumber
\label{eq:casdefs}
\eea
This is easily verified using the commutators:
\bea
[ W_m,Q_{\alpha} ] &=& -i\sigma _{mn\alpha}^{\beta}
Q_{\beta}P^n \quad , \nonumber\\
{[} \barQ _{\betadot} \barsigma _m^{\betadot\gamma}
Q_{\gamma},Q_{\alpha} {]} &=& -2P_mQ_{\alpha}
+4i\sigma_{nm\alpha}^{\beta}P^nQ_{\beta} \quad ,
\label{eq:handycoms}
\eea
which imply:
\bea
[ C_{mn},Q_{\alpha} ] &=& [B_m,Q_{\alpha}]P_n
- [B_n,Q_{\alpha}]P_m \nonumber\\
&=& 0 \qquad .
\label{eq:itsthecas}
\eea

\subsection{Classification of SUSY irreps on single particle states}

We now have enough machinery to construct all possible
irreducible representations of supersymmetry on asymptotic
(on-shell) physical states. We begin with N=1 SUSY, treating the
massive and massless states separately. Unlike the case of
\poin\ symmetry, we do not have to consider tachyons -- they
are forbidden by the fact that $\{ Q,\barQ \}$ is positive
definite.

\subsubsection{N=1 SUSY, massive states}

We analyze massive states from the rest frame
$P_m = (m,\vec{0})$. We can write:
\bea
C^2 &=& 2m^4J_iJ^i \quad , \nonumber\\
J_i &\equiv & S_i - {1\over 4m} \barQ
\barsigma _i Q \quad ,
\label{eq:jdefhere}
\eea
where $S_i$ is the spin operator and $i$ is a
spatial index: $i=1,2,3$.
Both $S_i$
and $\barsigma _i^{\alphadot\beta}$ obey the $SU(2)$
algebra, so
\be
[ J_i,J_j ] = i\epsilon _{ijk} J_k \quad ,
\label{eq:jissutwo}
\ee
and $J^2$ has eigenvalues $j(j+1)$, $j$ equal integers or
half-integers.

The commutator of $J_i$ with either $Q$ or $\barQ$
is proportional to $\vec{P}$ and thus vanishes since
we are in the rest frame. $Q_{\alpha}$, $\barQ _{\alphadot}$
are in fact two pairs of creation/annihilation operators
which fill out the N=1 massive SUSY irrep of fixed $m$ and $j$:
\be
\{ Q_{\alpha},\barQ _{\betadot} \} = 2m\sigma ^0_{\alpha\betadot}
= 2m\left( \mymatrix{1&0\cr
              0&1\cr} \right) \quad .
\label{eq:cliffalg}
\ee
Given any state of definite $\ket{m,j}$ we can define
a new state
\bea
\cliff &=& Q_1Q_2\ket{m,j} \quad ,\label{eq:defcliff}\\
Q_1\cliff &=& Q_2\cliff \quad =\quad  0 \quad . \nonumber
\eea
Thus $\cliff$ is a Clifford vacuum state with respect to
the fermionic annihilation operators $Q_1$, $Q_2$.
Note that $\cliff$ has degeneracy $2j$$+$$1$ since
$j_3$ takes values $-j,\ldots j$.

Acting on $\cliff$, $J_i$ reduces to just the
spin operator $S_i$, so $\cliff$ is actually an
eigenstate of spin:
\be
\cliff = \ket{m,s,s_3} \quad .
\label{eq:cliffspin}
\ee
Thus we can characterize all the states in the SUSY
irrep by mass and spin.

It is convenient to define conventionally normalized
creation/annihilation operators:
\bea
a_{1,2} &=& {1\over\sqrt{2m}}Q_{1,2} \quad , \nonumber\\
a^{\dagger}_{1,2} &=& {1\over\sqrt{2m}}
\barQ_{\onedot ,\twodot} \quad .
\label{eq:convennorm}
\eea
Then for a given $\cliff$ the full massive SUSY irrep is:
\bea
\cliff &&\nonumber\\
a^{\dagger}_1 \cliff &&\\
a^{\dagger}_2 \cliff &&\nonumber\\
{1\over\sqrt{2}}a^{\dagger}_1a^{\dagger}_2 \cliff &
=& -{1\over\sqrt{2}}a^{\dagger}_2a^{\dagger}_1\cliff \nonumber
\label{eq:thestateshere}
\eea
There are a total of $4(2j$$+$$1)$ states in the massive irrep.

We compute the spin of these states by using the commutators:
\be
\big[ S_3,\left( {a^{\dagger}_2\atop a^{\dagger}_1}
\right) \big]
= {1\over 2} \left( {a^{\dagger}_2\atop -a^{\dagger}_1} \right)
\quad .
\label{eq:spincomguy}
\ee
Thus for $\cliff = \ket{m,j,j_3}$ we get states of
spin $s_3 = j_3$, $j_3$$-$$\half$, $j_3$$+$$\half$, $j_3$.

As an example, consider the $j$$=$$0$ or {\bf fundamental}
N=1 massive irrep. Since $\cliff$ has spin zero there are a
total of four states in the irrep, with spins $s_3 = 0$,
$-\half$, $\half$, and $0$, respectively.
Since the parity operation interchanges $a^{\dagger}_1$ with
$a^{\dagger}_2$, one of the spin zero states is a pseudoscalar.
Thus these four states
correspond to one massive Weyl fermion, one real scalar,
and one real pseudoscalar.

\subsubsection{N=1 SUSY, massless states}

We analyze massless states from the light-like reference frame
$P_m = (E,0,0,E)$. In this case
\be
C^2 = -2E^2(B_0 - B_3)^2 = -\half E^2
\barQ _{\twodot}Q_2\barQ _{\twodot}Q_2 = 0 \quad .
\label{eq:ciszero}
\ee
Also we have:
\bea
\{ Q_{\onedot},\barQ _{\onedot} \} &=& 4E \quad ,\nonumber\\
\{ Q_{\twodot},\barQ _{\twodot} \} &=& 0 \quad .
\label{eq:masslessacom}
\eea
We can define a vacuum state $\cliff$ as in the massive case.
However we notice from Eq.~\ref{eq:masslessacom}
that the creation operator $\barQ _{\twodot}$
makes states of zero norm:
\be
\bra{\Omega} Q_2\barQ_{\twodot} \cliff = 0 \quad .
\label{eq:iszeronorm}
\ee
This means that we can set $\barQ _{\twodot}$ equal to
zero in the operator sense. Effectively there is just
one pair of creation/annihilation operators:
\be
a = {1\over 2\sqrt{E}}Q_1 \quad ,\quad
a^{\dagger} = {1\over 2\sqrt{E}}\barQ _{\onedot} \quad .
\label{eq:justonepair}
\ee
$\cliff$ is nondegenerate and has definite helicity $\lambda$.
The creation operator $a^{\dagger}$ transforms like
$(0,\half)$ under the Lorentz group, thus it increases helicity
by $1/2$. The massless N=1 SUSY irreps each contain two states:
\bea
\cliff &\; &{\rm helicity\ }\lambda \quad ,\nonumber\\
a^{\dagger} \cliff &\; &{\rm helicity\ }\lambda +\half \quad .
\label{eq:twostateshere}
\eea
However this is not a CPT eigenstate in general, requiring that
we pair two massless SUSY irreps to obtain four states with
helicities $\lambda$, $\lambda$$+$$\half$, $-\lambda$$-$$\half$,
and $-\lambda$.

\subsubsection{N$>$1 SUSY, no central charges, massless states}

Here we have N creation operators $a^{\dagger}_A$.
These generate a total of $2^N$ states in the SUSY irrep.
The states have the form:
\be
{1\over \sqrt{n!}}a^{\dagger}_{A_1}\ldots a^{\dagger}_{A_n}
\cliff \quad ,
\label{eq:typstate}
\ee
with degeneracy given by the binomial coefficient
$\left( {N\atop n} \right)$. Denoting the helicity of
$\cliff$ by $\lambda$, the helicities in the irrep are
$\lambda$, $\lambda$$+$$\half$, \ldots $\lambda$$+$$\Ntwo$.
This is not a CPT eigenstate except in the special
case $\lambda = -N/4$. Examples of some of the more
important irreps are given in Table~\ref{tab:masslessirrept}.

\begin{table}[t]
\caption{Examples of N$>$1 massless SUSY irreps (no central charge)
\label{tab:masslessirrept}}

\vspace{0.4cm}
\begin{center}
\begin{tabular}{|l|c|c|c|c|c|c|c|c|c|c|c|}
\hline
{\bf N=2}&\mco{11}{|l|}{ }\\ \hline
&&&&&\mco{7}{|l|}{ }\\
$\Omega _0$&
helicity&0&$\half$&1&\mco{7}{|l|}{ }\\
\cline{2-5}
&no. of states&1&2&1&
\mco{7}{|l|}{$\Omega _0$ and $\Omega _{-1}$ together}\\
\cline{1-5}
&&&&&\mco{7}{|l|}{make one N=2 on-shell}\\
$\Omega _{-1}$&
helicity&$-1$&$-\half$&0&\mco{7}{|l|}{vector multiplet.}\\
\cline{2-5}
&no. of states&1&2&1&\mco{7}{|l|}{ }\\
\hline
&&&&&\mco{7}{|l|}{ }\\
$\Omega _{-\half}$&
helicity&$-\half$&0&$\half$&\mco{7}{|l|}{A massless N=2 on-shell}\\
\cline{2-5}
&no. of states&1&2&1&\mco{7}{|l|}{hypermultiplet.}\\
\hline
{\bf N=4}&\mco{11}{|l|}{ }\\ \hline
&&&&&&&\mco{5}{|l|}{}\\
$\Omega _{-1}$&
helicity&$-1$&$-\half$&0&$\half$&1&\mco{5}{|l|}{A massless N=4 on-shell}\\
\cline{2-7}
&no. of states&1&4&6&4&1&\mco{5}{|l|}{vector multiplet.}\\
\hline
{\bf N=8}&\mco{11}{|l|}{ }\\ \hline
&&&&&&&&&&&\\
$\Omega _{-2}$&
helicity&$-2$&$-\threehalf$&$-1$&$-\half$&$0$&$
\half$&1&$\threehalf$&2&An N=8 gravity\\
\cline{2-11}
&no. of states&1&8&28&56&70&56&28&8&1&multiplet.\\
\hline
\end{tabular}
\end{center}
\end{table}

\subsubsection{N$>$1 SUSY, no central charges, massive states}

In this case we have 2N creation operators
$(a^A_{\alpha})^{\dagger}$. There are
$2^{2N}(2j+1)$ states in a massive irrep.
Consider, for example, the fundamental N=2 massive irrep:
$$\begin{array}{llccc}
\Omega_0:\quad&{\rm spin}&0&\half&1\\
&{\rm no.~of~spin~irreps}&5&4&1\\
&{\rm total~no.~of~states}&5&8&3
\end{array}$$
There are a grand total of 16 states. Let us describe them
in more detail:
$$\begin{array}{lrc}
{\rm 1~state:~}&\cliff &{\rm 1~spin~0~state}\\
{\rm 4~states:~}&(a^A_{\alpha})^{\dagger}\cliff &{\rm 4~spin~\half ~states}\\
{\rm 6~states:~}&(a^{A_1}_{\alpha _1})^{\dagger}
(a^{A_2}_{\alpha _2})^{\dagger}\cliff &
{\rm 3~spin~1~and~3~spin~0~states}\\
{\rm 4~states:~}&(a^{A_1}_{\alpha _1})^{\dagger}
(a^{A_2}_{\alpha _2})^{\dagger}(a^{A_3}_{\alpha _3})^{\dagger}
\cliff &{\rm 4~spin~\half ~states}\\
{\rm 1~state:~}&(a^{A_1}_{\alpha _1})^{\dagger}
(a^{A_2}_{\alpha _2})^{\dagger}(a^{A_3}_{\alpha _3})^{\dagger}
(a^{A_4}_{\alpha _4})^{\dagger}\cliff &
{\rm 1~spin~0~state}
\end{array}$$
The only counting which is not obvious is $6=3$ spin 1 $+3$ spin 0;
this can be verified by looking at the Lorentz group tensor
products:
\begin{eqnarray*}
\lefteqn{\big[ (0,\half )^1 \oplus (0,\half )^2 \big]
\otimes \big[ (0,\half )^1 \oplus (0,\half )^2 \big] = } \\
& & (0,0) + \big[ (0,1) + (0,0) \big] + (0,0) \quad .
\end{eqnarray*}
The key point is that
$(a^{A}_{\alpha})^{\dagger}(a^{A}_{\beta})^{\dagger}\cliff$,
by antisymmetry, only contains the singlet.

\subsubsection{N$>$1 SUSY, with central charges}

In the presence of central charges $\barQ _{\alphadot A}$,
$Q^A_{\alpha}$ cannot be interpreted in terms of
creation/annihilation operators without rediagonalizing
the basis. Recall
\bea
\{ Q_{\alpha}^A,Q_{\beta}^B \} &=& \epsd X^{AB} \quad ,\nonumber\\
\{ \barQ_{\alphadot A},\barQ_{\betadot B} \} &=&
-\epsilon _{\alphadot\betadot} X^*_{AB} \quad ,
\label{eq:nonzeroccqq}
\eea
where $X^{AB}$ is antisymmetric and, following Wess and Bagger,
we impose the convention $X^{AB}$$=$$-X_{AB}$.

Since the central charges commute with all the other generators,
we can choose any convenient basis to describe them.
We will use Zumino's decomposition of a general complex
antisymmetric matrix: \cite{zumino}
\be
X^{AB} = U^A_C\tilde{X}^{CD}(U^T)_D^B \quad ,
\label{eq:zurel}
\ee
where, for N even, $\tilde{X}^{CD}$ has the form
\be
\left( \mymatrix{(Z_1\epsilon ^{ab})&0&\ldots &0\cr
                    0&(Z_2\epsilon ^{ab})&\ldots &0\cr
                    \vdots &\vdots &\ddots &\vdots\cr
                    0&0& \ldots & (Z_{{N\over 2}}\epsilon ^{ab})\cr}
\right)\quad ,
\label{eq:zumatrix}
\ee
where $\epsilon ^{ab}$$=$$i\sigma ^2$.
For N odd, there is an extra right-hand column of zeroes
and bottom row of zeroes. In this decomposition the
``eigenvalues''
$Z_1$, $Z_2$, $\ldots$ $Z_{\left[{N\over 2}\right]}$ are
real and nonnegative.

Consider now the massive states in the rest frame.
In the basis defined by Zumino's decomposition we have:
\bea
\{ Q_{\alpha},\barQ _{\betadot} \} &=&
2m\sigma ^0_{\alpha\betadot}\delta ^a_b \delta ^L_M
\quad ,\nonumber\\
\{ Q_{\alpha}^{aL},Q_{\beta}^{bM} \} &=&
\epsd \epsilon ^{ab}\delta ^{LM}Z_M \quad ,\label{eq:specbasacc}\\
\{ \barQ_{\alphadot aL},\barQ_{\betadot bM} \} &=&
-\epsilon _{\alphadot\betadot} \epsilon _{ab}
\delta _{LM}Z_M \quad ,\nonumber
\eea
where the internal indices $A$, $B$ have now been
replaced by the index pairs $(a,L)$, $(b,M)$, with
$a,b = 1,2$ and $L,M = 1,2,\ldots [\Ntwo ]$.
Here and in the following, the repeated $M$ index
is {\bf not} summed over.

It is now apparent that there are 2N pairs
of creation/annihilation operators:
\bea
a_{\alpha}^L &=& {1\over\sqrt{2}}
\left[ Q_{\alpha}^{1L} + \epsd
\barQ _{\gammadot 2L}\barsigma ^{0\gammadot\beta} \right]
\quad ,\nonumber\\
(a_{\alpha}^L)^{\dagger} &=& {1\over\sqrt{2}}
\left[ \barQ _{\alphadot 1L} + \epsilon _{\alphadot\betadot}
\barsigma ^{0\betadot\gamma}Q_{\gamma}^{2L} \right]
\quad ,\nonumber\\
b_{\alpha}^L &=& {1\over\sqrt{2}}
\left[ Q_{\alpha}^{1L} - \epsd
\barQ _{\gammadot 2L}\barsigma ^{0\gammadot\beta} \right]
\quad ,\label{eq:newcreops}\\
(b_{\alpha}^L)^{\dagger} &=& {1\over\sqrt{2}}
\left[ \barQ _{\alphadot 1L} - \epsilon _{\alphadot\betadot}
\barsigma ^{0\betadot\gamma}Q_{\gamma}^{2L} \right]
\quad .\nonumber
\eea
The Lorentz index structure here looks a little strange, but
the important point is that $\barQ_{\alphadot}$ transforms
the same as $Q^{\alpha}$ under {\bf spatial} rotations. Thus
$(a_{\alpha}^L)^{\dagger}$, $(b_{\alpha}^L)^{\dagger}$ create
states of definite spin.

The anticommutation relation are:
\bea
\{ a_{\alpha}^L,(a_{\beta}^M)^{\dagger} \} &=&
(2m+Z_M)\sigma ^0_{\alpha\betadot}\delta ^L_M \quad ,\nonumber\\
\{ b_{\alpha}^L,(b_{\beta}^M)^{\dagger} \} &=&
(2m-Z_M)\sigma ^0_{\alpha\betadot}\delta ^L_M \quad .
\label{eq:coolanticomms}
\eea
This is easily verified from
Eqs~\ref{eq:specbasacc}, \ref{eq:newcreops} using the
relations:
\bea
\epsilon _{\alpha\delta}
\barsigma ^{0\gammadot\delta} \epsilon _{\gammadot\betadot}
&=& -\sigma ^0_{\alpha\betadot} \quad ,\nonumber\\
\epsilon _{\betadot\deltadot} \barsigma ^{0\deltadot\gamma}
\epsilon _{\alpha\gamma} &=& \sigma ^0_{\alpha\betadot}
\quad .
\label{eq:coolidents}
\eea

\subsubsection{BPS-saturated states}

Since $\{ a,a^{\dagger} \}$ and $\{ b,b^{\dagger} \}$
are positive definite operators, and since the $Z_M$
are nonnegative, we deduce the following:
\begin{itemize}
\item For all $Z_M$ in any SUSY irrep:
\be
Z_M \le 2m \qquad .
\label{eq:ccinequal}
\ee
\item When $Z_M < 2m$ the multiplicities of the massive
irreps are the same as for the case of no central charges.
\item The special case is when we saturate the bound,
i.e. $Z_M = 2m$ for some or all $Z_M$. If e.g. all the
$Z_M$ saturate the bound, then all of the $(b_{\alpha}^L)^{\dagger}$
are projections onto zero norm states; thus effectively we
lose half of the creation operators. This implies that
this massive SUSY irrep has only $2^N(2j$$+$$1)$ states
instead of $2^{2N}(2j$$+$$1)$ states.
\end{itemize}

These reduced multiplicity massive multiplets are often
called {\bf short multiplets}. The states are often referred
to as {\bf BPS-saturated states}, because of the connection
to BPS monopoles in supersymmetric gauge theories. \cite{bps}

For example, let us compare the fundamental N=2 massive
irreps. For N=2 there is only one central charge, $Z$.
For $Z < 2m$ we have the {\bf long multiplet}
already discussed:
$$\begin{array}{llccc}
\Omega_0^{\rm long}:\quad&{\rm spin}&0&\half&1\\
&{\rm no.~of~spin~irreps}&5&4&1\\
&{\rm total~no.~of~states}&5&8&3
\end{array}$$
There are a grand total of 16 states.

For $Z = 2m$ we have BPS-saturated states in a
short multiplet:
$$\begin{array}{llccc}
\Omega_0^{\rm short}:\quad&{\rm spin}&0&\half&1\\
&{\rm no.~of~spin~irreps}&2&1&0\\
&{\rm total~no.~of~states}&2&2&0
\end{array}$$
There are a grand total of 4 states. Note that
the spins and number of states of this BPS-saturated
massive multiplet match those of the N=2 {\bf massless}
hypermultiplet in Table~\ref{tab:masslessirrept}.

Let us also compare the $j$$=$$\half$ N=2 massive irreps.
For $Z < 2m$ we have a long multiplet with 32 states:
$$\begin{array}{llcccc}
\Omega_{\half}^{\rm long}:\quad&{\rm spin}&0&\half&1&\threehalf\\
&{\rm no.~of~spin~irreps}&4&6&4&1\\
&{\rm total~no.~of~states}&4&12&12&4
\end{array}$$
For $Z = 2m$ we have a short multiplet with 8 states:
$$\begin{array}{llcccc}
\Omega_{\half}^{\rm short}:\quad&{\rm spin}&0&\half&1&\threehalf\\
&{\rm no.~of~spin~irreps}&1&2&1&0\\
&{\rm total~no.~of~states}&1&4&3&0
\end{array}$$
Note that the spins and number of states of this
BPS-saturated massive multiplet match those of the
N=2 massless vector multiplet (allowing for the fact
that a massless vector eats a scalar in becoming massive).

\subsubsection{Automorphisms of the supersymmetry algebra}
In the absence of central charges, the general 4-dimensional
SUSY algebra has an obvious $U(N)$ automorphism symmetry:
\be
Q_{\alpha}^A \rightarrow \hbox{$U^A$}_B Q^B_{\alpha},\qquad
\barQ_{\alphadot A} \rightarrow \barQ_{\alphadot B}
\hbox{$U^{\dagger B}$}_A \quad ,
\ee
where \hbox{$U^A$}$_B$ is a unitary matrix. SUSY irreps
on asymptotic single particle states will
automatically carry a representation of the automorphism group.
For massless irreps $U(N)$ is the largest automorphism symmetry
which respects helicity.

For massive irreps, we have already noted that $Q^{\alpha}$ and
$\barQ_{\alphadot}$ transform the same way under spatial rotations.
Assembling these into a 2N component object, one finds that the
largest automorphism group which respects spin is $USp(2N)$,
the unitary symplectic group of rank N. \cite{fsz}
In the presence of central charges, the automorphism group is
still $USp(2N)$ provided that none of the central charges
saturates the BPS bound; this follows from our ability to make the
basis change Eqs.~\ref{eq:newcreops}, \ref{eq:coolanticomms}.
When one central charge saturates the BPS bound, the automorphism
group is reduced to $USp(N)$ for N even, $USp(N+1)$ for N odd.

The automorphism symmetries give us constraints on the internal
symmetry group generated by the $B_{\ell}$. In the case of
no central charges $U(N)$ is the largest possible internal
symmetry group which can act nontrivially on the $Q$'s.
With a single central charge, the intertwining relation
Eq.~\ref{eq:intertwinerel} implies that $USp(N)$ is the
largest such group.

\subsubsection{Supersymmetry represented on quantum fields}

So far we have only discussed representations of SUSY
on asymptotic states, not on quantum fields.
$Q_{\alpha}$, $\barQ _{\alphadot}$ can be represented as
superspace differential operators acting on fields.
The Clifford vacuum condition Eq.~\ref{eq:defcliff}
becomes a commutation condition:
\be
[ \barQ _{\alphadot},\Omega (x) ] = 0 \quad .
\label{eq:cliffcomcon}
\ee

For {\bf on-shell} fields,
the construction of SUSY irreps proceeds as before,
with the following exception. If $\Omega (x)$
is a {\bf real} scalar field, then the adjoint of
Eq.~\ref{eq:cliffcomcon} is
\be
[ Q _{\alpha},\Omega (x) ] = 0 \quad .
\label{eq:cliffcomadj}
\ee
In that case, Eqs.~\ref{eq:cliffcomcon}, \ref{eq:cliffcomadj}
together with the Jacobi identity for
$\{ [ \Omega (x),Q],\barQ \}$ implies that $\Omega (x)$
is a constant.

Thus we conclude that $\Omega (x)$ must be a {\bf complex}
scalar field. This has the effect that some SUSY on-shell irreps
on fields have twice as many field components as the corresponding irreps
for on-shell states. Because we already paired up most
SUSY irreps on states to get CPT eigenstates, this doubling
really only effects the SUSY irreps based on the special
case $\Omega _{\lambda}$, $\lambda = -N/4$. The first example
is the massless N=2 hypermultiplet. On asymptotic single particle
states this irrep consists of 4 states
(see Table~\ref{tab:masslessirrept}); the massless N=2
hypermultiplet on fields, however, has 8 real components.

\subsection{N=1 rigid superspace}

Relativistic quantum field theory relies upon the fact
that the spacetime coordinates $x^m$ parametrize the
coset space defined as the \poin\ group
modded out by the Lorentz group.
Clearly it is desirable to find a similar coordinatization
for supersymmetric field theory. For simplicity we will
discuss the case of N=1 SUSY, deferring N$>$1 SUSY until
Section 5.

The first step is to rewrite the N=1 SUSY algebra as a Lie algebra.
This requires that we introduce constant Grassmann spinors
$\tha$, $\bartha$:
\be
\{ \tha ,\th ^{\beta} \} = \{ \bartha ,\barth _{\betadot} \} =
\{ \tha ,\barth _{\betadot} \} = 0 \quad .
\label{eq:grassdefs}
\ee
This allows us to replace the anticommutators in the
N=1 SUSY algebra with commutators:
\bea
{[}\, \th\, Q, \barth\,\barQ \,{]} &=& 2\th \sigma ^m\barth P_m
\quad ,\nonumber\\
{[}\, \th\, Q, \th\, Q \,{]} &=& 0 \quad ,\\
{[}\, \barth\,\barQ ,\barth\,\barQ \,{]} &=& 0 \quad .\nonumber
\label{eq:antitocom}
\eea
Note we have now begun to employ the spinor summation convention
discussed in the Appendix.

Given a Lie algebra we can exponentiate to get the general
group element:
\vskip .1in
\be
G(x,\th ,\barth ,\omega ) =
\e{i[-x^mP_m + \th \,Q + \barth\,\barQ ]}
\e{-{i\over 2}\omega ^{mn}M_{mn}}
\quad ,
\label{eq:gengroupel}
\ee
\vskip .1in
\noindent where the minus sign in front of $x^m$ is a
convention. Note that this form of the
general N=1 super\poin\ group element is unitary
since $(\th Q)^{\dagger}$$=$$\barth\,\barQ$.

{}From Eq.~\ref{eq:gengroupel} it is clear that
$(x^m,\tha ,\bartha )$ parametrizes a $4$$+$$4$
dimensional coset space: N=1 super\poin\ {\bf mod} Lorentz.
This coset space is more commonly known as
{\bf N=1 rigid superspace}; ``rigid'' refers to the
fact that we are discussing {\bf global} supersymmetry.

There are great advantages to constructing supersymmetric
field theories in the superspace/superfield formalism,
just as there are great advantages to constructing
relativistic quantum field theories in a manifestly Lorentz
covariant formalism. Our rather long technical detour
into superspace and superfield constructions will pay off
nicely when we begin the construction of supersymmetric actions.

\subsubsection{Superspace derivatives}

Here we collect the basic notation and properties of
N=1 superspace derivatives.

\bea
\dd &=& {\textstyle\partial\over\partial\tha}\quad ,\;\quad
\du = {\textstyle\partial\over\partial\th _{\alpha}}
= -\epsu \partial _{\beta} \quad ,\nonumber\\
\bardu &=& {\textstyle \partial\over\partial\bartha}\quad ,\;\quad
\bardd = {\textstyle \partial\over\partial\barth ^{\alphadot}}
= -\epsilon _{\alphadot\betadot}
\overline{\partial}^{\betadot}\quad ,\nonumber\\
\dd \theta ^{\beta} &=& \delta _{\alpha}^{\beta}\quad ,\;\quad\quad
\bardu \barth _{\betadot} = \delta _{\betadot}^{\alphadot}
\quad ,\nonumber\\
\du\th ^{\beta} &=& -\epsu  \quad ,\quad
\dd\th _{\beta} = -\epsd \quad ,\\
\bardu\barth ^{\betadot} &=& -\epsilon ^{\alphadot\betadot}\quad ,\quad
\bardd\barth _{\betadot} = -\epsilon _{\alphadot\betadot}\quad ,\nonumber\\
\dd \th ^{\beta}\th ^{\gamma} &=&
\delta _{\alpha}^{\beta}\th ^{\gamma}
-\delta _{\alpha}^{\gamma}\th ^{\beta}\quad ,\nonumber\\
\dd (\th\th ) &=& 2\th _{\alpha} \quad ,\quad
\;\;\bardd (\barth\barth ) = -2\bartha \quad ,\nonumber\\
\partial ^2 (\th\th ) &=& 4\quad ,\quad
\quad\;\;\bard ^2 (\barth\barth ) = 4\quad .\nonumber
\label{eq:supdrels}
\eea

\subsubsection{Superspace integration}

We begin with the Berezin integral for a single Grassmann
parameter $\th$:
\bea
\int d\th\; \th &= 1 \qquad ,\nonumber\\
\int d\th &= 0 \qquad ,
\label{eq:berezin}\\
\int d\th\; f(\th ) &= f_1 \qquad ,\nonumber
\eea
where we have used the fact that an arbitrary function
of a single Grassmann parameter $\th$ has the Taylor
series expansion $f(\th ) = f_0 + \th f_1$.

We note three facts which follow from the definitions
of Eq.~\ref{eq:berezin}.
\begin{itemize}
\item Berezin integration is translationally invariant:
\bea
\int d(\th +\xi )\; f(\th +\xi ) &=& \int d\th\; f(\th )
\quad ,\nonumber\\
\int d\th\; {d{\rm\ }\over d\th}f(\th ) &=& 0 \quad .
\label{eq:transint}
\eea
\item Berezin integration is equivalent to differentiation:
\be
{d{\rm\ }\over d\th}f(\th ) = f_1 =
\int d\th\; f(\th )\quad .
\label{eq:intisdiff}
\ee
\item We can define a Grassmann delta function by
\be
\delta (\th ) \equiv \th \qquad .
\label{eq:deltadef}
\ee
\end{itemize}

These results are easily generalized to the case
of the N=1 superspace coordinates $\tha$, $\bartha$.
The important notational conventions are:
\bea
d^2\th &=& -\quarter d\tha\, d\th ^{\beta}\, \epsd
\quad ,\nonumber\\
d^2\barth &=& -\quarter d\bartha\, d\barth _{\betadot}\,
\epsilon ^{\alphadot\betadot} \quad ,\\
d^4\th &=& d^2\th\; d^2\barth \quad .\nonumber
\label{eq:intsuperdefs}
\eea
Using this notation and the spinor summation convention,
we have the following identities:
\bea
\int d^2\th\; \th\,\th &=& 1\qquad ,\nonumber\\
\int d^2\barth\; \barth\,\barth &=& 1\qquad .
\label{eq:intidents}
\eea

\subsubsection{Superspace covariant derivatives}

If we wanted to treat a general curved N=1 superspace,
we would have to introduce a $4$$+$$4$$=$$8$-dimensional
vielbein and spin connection. Using $M$ to denote an
8-dimensional superspace index, and $A$ to denote
an 8-dimensional super-tangent space index, we can
write the vielbein and spin connection as
$E_M^A$ and $W_A^{mn}$ respectively. The general form
of a covariant derivative in such a space is thus
\be
D_M = E_M^A(\partial _A + \half W_A^{mn}M_{mn}) \quad ,
\label{eq:gencovderdef}
\ee
where $\partial _A$$=$$(\partial _m,\dd ,\bardd )$.

Naively one might expect that $D_M$ reduces to $\partial _M$
for N=1 rigid superspace, since the rigid superspace has
zero curvature. However it is possible to show \cite{west}
that N=1 rigid superspace has nonzero {\bf torsion}, and
thus that the vielbein is nontrivial.
The covariant derivatives for N=1 rigid superspace are given by:
\bea
D_m &=& \partial _m \qquad ,\nonumber\\
\D &=& \dd + i\sigma ^m_{\alpha\betadot}
\barth ^{\betadot}\partial _m \quad ,\nonumber\\
\Ddot &=& -\bardd - i\th ^{\beta}\sigma ^m_{\beta\alphadot}
\partial _m \quad ,\\
D^{\alpha} &=& -\du - i\barth _{\betadot}
\barsigma ^{m\betadot\alpha}\partial _m\quad ,\nonumber\\
\barD ^{\alphadot} &=& \bardu + i\barsigma ^m_{\alphadot\beta}
\th _{\beta}\partial _m \quad .\nonumber
\label{eq:covderdef}
\eea

\section{N=1 Superfields}

\subsection{The general N=1 scalar superfield}

The general scalar superfield $\Phi\sx$ is just
a scalar function in N=1 rigid superspace. It has
a finite Taylor expansion in powers of $\tha$, $\bartha$;
this is known as the {\bf component expansion} of the
superfield:
\bea
\lefteqn{\Phi\sx = f(x) + \th\phi (x) + \barth\barchi (x)
+\th\th m(x) +\barth\barth n(x) } \nonumber \\
&&+ \th\sigma ^m\barth v_m(x) + (\th\th )\barth\barlam (x) +
(\barth\barth )\th\psi (x) + (\th\th )(\barth\barth )d(x)
\quad .
\label{eq:compgenscal}
\eea
The component fields in Eq~\ref{eq:compgenscal} are complex;
redundant terms like $\barth\barsigma ^m\th v_m$ have already
been removed using the Fierz identities listed in the Appendix.
The fermionic component fields $\phi (x)$, $\barchi (x)$,
$\barlam (x)$, and $\psi (x)$ are Grassmann odd, i.e. they
anticommute with each other and with $\th$, $\barth$.

To compute the effect of an infinitesimal N=1 SUSY transformation
on a general scalar superfield, we need the explicit representation
of $Q$, $\barQ$ as superspace differential operators.
Recall that for ordinary scalar fields the translation generator
$P_m$ is represented (with our conventions)
by the differential operator $i\partial _m$.
Let $\xi ^{\alpha}$ be a constant Grassmann complex
Weyl spinor, and
consider the effect of left multiplication by a
``supertranslation'' generator $G(y,\xi )$ on an
arbitrary coset element $\Omega\sx$:
\bea
\lefteqn{ G(y,\xi )\Omega\sx = \e{i{[}-y^mP_m+\xi Q+\barxi\barQ {]}}
\e{i{[}-x^mP_m+\th Q +\barth\barQ {]} }}\nonumber\\
&&=\e{i[-(x^m +y^m)P_m + (\tha + \xi ^{\alpha})Q_{\alpha}
+(\bartha + \barxi _{\alphadot} )\barQ ^{\alphadot}
+{i\over 2}[\xi Q,\barth\barQ ] + {i\over 2}[
\barxi\barQ ,\th Q]} \nonumber\\
&&= \Omega\left( (x^m + y^m - i\xi\sigma ^m\barth +
i\th\sigma ^m\barxi ), \th +\xi , \barth + \barxi \right) \quad ,
\label{eq:supertran}
\eea
where, to obtain the last expression, we have used the
commutators:
\bea
{[} \xi \,Q,\barth\,\barQ {]} &=& 2\xi\sigma ^m\barth P_m
\quad ,\nonumber\\
{[} \barxi\,\barQ ,\th\, Q {]} &=& -2\th\sigma ^m\barxi P_m \quad .
\label{eq:justsomecoms}
\eea
{}From Eq.~\ref{eq:supertran} we see that, with our conventions,
$P_m$, $Q$, and $\barQ$ have the following representation
as superspace differential operators:
\bea
P_m &:& i\partial _m \quad ,\nonumber\\
Q_{\alpha} &:& \dd - i\sigma ^m_{\alpha\betadot}\barth ^{\betadot}
\partial _m \quad ,\\
\barQ_{\alphadot} &:& \bardd - i\th ^{\beta}\sigma ^m_{\beta\alphadot}
\partial _m \quad .\nonumber
\label{eq:qsasdiffs}
\eea

It is now a trivial matter to compute the infinitesimal
variation of the general scalar superfield Eq.\ref{eq:compgenscal}
under an N=1 SUSY transformation:
\bea
\lefteqn{\dx\Phi\sx = (\xi Q + \barxi\barQ )\Phi\sx}\nonumber\\
&&= \xi\phi + \barxi\barchi + i\th\sigma ^m\barxi\partial _mf
+ 2\xi\th m +\th\sigma ^m\barxi v_m -i\xi\sigma ^m\barth\partial _mf\nonumber\\
&& + 2\barxi\barth n
+\xi\sigma ^m\barth v_m + i(\th\sigma ^m\barxi )\th\partial _m\phi
+(\th\th )(\barxi\barlam ) -i(\xi\sigma ^m\barth )\barth\partial _m\barchi\\
&& +
(\barth\barth )(\xi\psi ) - i\xi\sigma ^m\barth\th\partial _m\phi
+ i\th\sigma ^m\barxi\barth\partial _m\barchi
+2(\xi\th )(\barth\barlam ) + 2(\barxi\barth )(\th\psi )\nonumber\\
&&-i\xi\sigma ^m\barth (\th\th )\partial _mm +
i\th\sigma ^m\barxi\th\sigma ^n\barth\partial _mv_n
+2(\th\th )(\barxi\barth )d
+i\th\sigma ^m\barxi (\barth\barth )\partial _mn
\nonumber\\
&&-i\xi\sigma ^m\barth\th\sigma ^n\barth\partial _mv_n
+2(\xi\th )(\barth\barth )d
-i\xi\sigma ^m\barth (\th\th )\barth\partial _m\barlam
+i\th\sigma ^m\barxi (\barth\barth )\th\partial _m\psi
\quad .\nonumber
\label{eq:biglongtransf}
\eea
Using the Fierz identities, we then have that
the component fields of
$\Phi$ transform as follows:
\bea
\dx f &=& \xi\phi + \barxi\barchi \quad ,\nonumber\\
\dx \phi _{\alpha} &=& 2\xi _{\alpha}m + \sigma ^m_{\alpha\betadot}
\barxi ^{\betadot}{[} i\partial _mf + v_m {]} \quad ,\nonumber\\
\dx \barchi ^{\alphadot} &=& 2\barxi ^{\alphadot}n +
\xi ^{\beta}\sigma ^m_{\beta\gammadot}\epsilon ^{\gammadot\alphadot}
{[} i\partial _mf - v_m {]}\quad ,\nonumber\\
\dx m &=& \barxi\barlam - {i\over 2}\partial _m\phi\sigma ^m\barxi
\quad ,\nonumber\\
\dx n &=& \xi\psi + {i\over 2}\xi\sigma ^m\partial _m\barchi \quad ,
\label{eq:stgenscal}\\
\dx v_m &=& \xi\sigma _m\barlam + \psi\sigma _m\barxi +
{i\over 2}\xi\partial _m\phi - {i\over 2}\partial _m\barchi\barxi
\quad ,\nonumber\\
\dx \barlam ^{\alphadot} &=& 2\barxi ^{\alphadot} d
+ {i\over 2}\barxi ^{\alphadot}\partial ^mv_m
+ i(\xi\sigma ^m\epsilon )^{\alphadot}\partial _m m
\quad ,\nonumber\\
\dx \psi _{\alpha} &=& 2\xi _{\alpha} d -
{i\over 2}\xi _{\alpha}\partial ^mv_m + i(\sigma ^m\barxi )_{\alpha}
\partial _m n \quad ,\nonumber\\
\dx d &=& {i\over 2}\partial _m \left[
\psi\sigma ^m\barxi + \xi\sigma ^m\barlam \right] \quad .\nonumber
\eea
Note the important fact that the complex scalar component field
$d(x)$ transforms by a total derivative.

We have thus demonstrated that the general scalar superfield
forms a basis for an (off-shell) linear representation of N=1
supersymmetry. However this representation is {\bf reducible}.
To see this, suppose we impose the following constraints on
the component fields of $\Phi\sf$:
\bea
\chi (x) &=& 0 \quad ,\nonumber\\
n(x) &=& 0 \quad ,\nonumber\\
v_m(x) &=& i\partial _m f(x) \quad ,
\label{eq:chirsubrel}\\
\barlam (x) &=& {i\over 2}\partial _m\phi\sigma ^m
\quad ,\nonumber\\
\psi (x) &=& 0 \quad ,\nonumber\\
d(x) &=& -\quarter \BBox f(x) \quad .\nonumber
\eea
It is easy to verify that the N=1 SUSY component field
transformations Eq.~\ref{eq:stgenscal} respect these
constraints. Thus the constrained superfield by itself
defines an off-shell linear representation of N=1 SUSY
(in fact, an irreducible representation). This suffices to
prove that representation defined by $\Phi\sf$ is reducible.
In fact there are several ways of extracting irreps
by constraining $\Phi\sf$, however the
general scalar superfield is {\bf not fully reducible},
i.e. the reducible representation is not a direct sum of
irreducible representations.

We can also use $\Phi\sf$ to demonstrate the importance
of the superspace covariant derivatives $\D$, $\Ddot$.
Consider $\bardd\Phi\sx$: this has fewer component fields
than $\Phi\sf$ since, for example, there is no
$(\th\th )(\barth\barth )$ term in its component expansion.
However the commutator of $\bardd$ with $\xi Q$ is nonvanishing:
\be
{[}\, \bardd ,\xi\, Q \,{]} = i\xi ^{\beta}\sigma ^m_{\beta\alphadot}
\partial _m \quad ,
\label{eq:nonzcomQ}
\ee
and this implies that an N=1 SUSY transformation generates
a $(\th\th )(\barth\barth )$ term. Thus $\bardd\Phi\sx$ is not
a true superfield in the sense of providing a basis for a
linear representation of supersymmetry.

The superspace covariant derivatives, on the other hand,
anticommute with $Q$ and $\barQ$:
\bea
\{ \D ,Q_{\beta} \} &= \{ \D ,\barQ _{\betadot} \} &= 0
\quad ,\nonumber\\
\{ \Ddot ,Q_{\beta} \} &= \{ \Ddot ,\barQ _{\betadot} \} &= 0
\quad ,
\label{eq:DQcoms}
\eea
Thus if $\Phi\sf$ is a general scalar superfield, then
$\partial _m\Phi$, $\D\Phi$, and $\Ddot\Phi$ are also
superfields.

\subsection{N=1 chiral superfields}
An N=1 chiral superfield is obtained by the constraints
Eq.~\ref{eq:chirsubrel} imposed on a general scalar superfield.
A more elegant and useful definition comes from
realizing that Eq.~\ref{eq:chirsubrel} is equivalent to
the following {\bf covariant constraint}:
\be
\Ddot\Phi\sf = 0 \qquad .
\label{eq:chsupdef}
\ee
Covariant constraints are constraints which involve
only superfields (and covariant derivatives of superfields,
since these are also superfields).
It is a plausible but nonobvious fact that the superfields
which define irreducible off-shell linear representations of
supersymmetry can always be obtained by imposing covariant
constraints on unconstrained superfields.

Let us find the most general solution to the covariant
constraint Eq.~\ref{eq:chsupdef}. Define new bosonic coordinates
$y^m$ in N=1 rigid superspace:
\be
y^m = x^m + i\th\sigma ^m\barth \quad .
\label{eq:ysdef}
\ee
We note in passing that the funny minus sign convention
in Eq.~\ref{eq:gengroupel} is tied the fact that sign in
Eq.~\ref{eq:ysdef} above is plus.
Since
\bea
\Ddot \,y^m &=& 0 \quad ,\nonumber\\
\Ddot \,\tha &=& 0 \quad ,
\label{eq:obviousders}
\eea
it is clear that any function $\Phi\sy$ of $y^m$ and $\tha$
(but not $\bartha$) satisfies
\be
\Ddot \,\Phi\sy = 0 \qquad .
\label{eq:itschiman}
\ee
It is easy to see that, since $\Ddot$ obeys the chain rule,
this is not just a particular solution of Eq.~\ref{eq:chsupdef}
but is in fact the most general solution.

Thus we may write the most general N=1 chiral superfield
as:
\be
\Phi\sy = A(y) + \sqrt{2}\th\psi (y) +\th\th F(y) \quad ,
\label{eq:mostgenchone}
\ee
where $A(y)$, $F(y)$ are complex scalar fields, while
$\psi ^{\alpha}(y)$ is a complex left-handed Weyl spinor.
The $\sqrt{2}$ is a convention. There are
$4$$+$$4 =8$ real off-shell field components; this is twice
the number in the on-shell fundamental N=1 massive irrep.

The full $\th$, $\barth$ component expansion is obtained
by using the Fierz identity Eq.~\ref{eq:goodfierz}. The result is:
\bea
\lefteqn{\Phi\sy = A(x) + \sqrt{2}\th\psi (x) +
\th\th F(x)} \nonumber\\
&&+i\th\sigma ^m\barth\partial _mA(x) +
{i\over\sqrt{2}}(\th\th )\partial _m\psi (x)\sigma ^m\barth -
\quarter (\th\th )(\barth\barth )\BBox A(x) \quad .
\label{eq:fullchiexp}
\eea
An infinitesimal N=1 SUSY transformation on the chiral
superfield yields:
\bea
\delta A &=& \sqrt{2}\xi\psi \quad ,\nonumber\\
\delta \psi &=& \sqrt{2}\xi F + \sqrt{2}i\sigma ^m\barxi
\partial _m A \quad ,\\
\delta F &=& -\sqrt{2}i\partial _m\psi\sigma ^m\barxi \quad .\nonumber
\eea
Note that $\delta F(x)$ is a total derivative.

Antichiral superfields, i.e. right-handed chiral superfields,
are defined in the obvious way. In particular, if $\Phi\sy$
is a chiral superfield, then $\Phi ^{\dagger}$ is an antichiral
superfield; it satisfies
\bea
\D\Phi ^{\dagger} &=& 0 \quad ,\nonumber\\
\Phi ^{\dagger} &=& \Phi ^{\dagger}(y^{\dagger},\barth );
\qquad y^{\dagger} = x^m - i\th\sigma ^m\barth \quad .
\eea

Since $\D$ and $\Ddot$ obey the chain rule, any product of
chiral superfields is also a chiral superfield, while any
product of antichiral superfields is also an antichiral
superfield. However it is also clear that if $\Phi\sy$ is
a chiral superfield, the following are {\bf not} chiral
superfields:
\begin{eqnarray*}
&&\Phi ^{\dagger}\Phi \quad ,\\
&&\Phi + \Phi ^{\dagger}\quad .
\end{eqnarray*}

For future reference, let us write down the
expressions for the covariant derivatives acting
on functions of $(y,\th ,\barth )$:

\bea
\D &=& \dd + 2i\sigma ^m_{\alpha\betadot}
\barth ^{\betadot}\partial _m \quad ,\nonumber\\
\Ddot &=& -\bardd \quad ,\nonumber\\
D^{\alpha} &=& -\du - 2i\barth _{\betadot}
\barsigma ^{m\betadot\alpha}\partial _m\quad ,\\
\barD ^{\alphadot} &=& \bardu \quad ,\nonumber
\label{eq:ycovderdef}
\eea
where of course here $\partial _m$ is a partial
derivative with respect to $y^m$ rather than $x^m$.

\subsection{N=1 vector superfields}

Vector superfields are defined from the general scalar
superfield by imposing a covariant reality constraint:
\be
V\sx = V^{\dagger}\sx \quad ,
\ee
or, in components:
\bea
f &=& f^* \quad ,\nonumber\\
\barchi &=& \phi ^* \quad ,\nonumber\\
m &=& n^* \quad ,\nonumber\\
v_m &=& v_m^* \quad ,\\
\barlam &=& \psi ^* \quad ,\nonumber\\
d &=& d^* \quad .\nonumber
\eea
Thus in components we have 4 real scalars, 2 complex Weyl spinors
(equivalently, 2 Majorana spinors), and 1 real vector. The
$8$$+$$8 = 16$ real components in this off-shell irrep are twice
the number in the on-shell $\Omega _{\half}$ massive irrep.

The presence of a real vector field in the N=1 vector
multiplet suggests we use vector superfields to construct
supersymmetric gauge theories. But first we must deduce the
superfield generalization of gauge transformations.

\subsubsection{Wess-Zumino gauge}

If $\Phi\sy$ is a chiral superfield, then $\Phi + \Phi ^{\dagger}$
is a special case of a vector superfield. In components:
\bea
\lefteqn{\Phi + \Phi ^{\dagger} = (A + A^*)
+ \sqrt{2}\th\psi + \sqrt{2}\barth\barpsi
+ \th\th F + \barth\barth F^*
+ i\th\sigma ^m\barth\partial _m(A-A^*)} \nonumber\\
&& +{i\over\sqrt{2}}(\th\th )\barth\barsigma ^m\partial _m\psi
+{i\over\sqrt{2}}(\barth\barth )\th\sigma ^m\partial _m\barpsi
-{\quarter}(\th\th )(\barth\barth )\BBox (A+A^*) \quad .
\label{eq:compphi}
\eea
{}From this we see that we can define the superfield analog
of an infinitesimal abelian gauge transformation to be
\be
V \rightarrow V + \Phi + \Phi ^{\dagger} \quad ,
\label{eq:abelgt}
\ee
since this definition gives the correct
infinitesimal transformation
for the vector component:
\bea
v_m &\rightarrow& v_m + \partial _m\Lambda \quad ;\\
\Lambda &=& i(A-A^*) \quad .\nonumber
\eea

The meaning of the ``bigger'' superfield transformation
Eq.~\ref{eq:abelgt} is that any superfield action invariant under
abelian gauge transformations will also be independent of
several component fields of $V\sx$. More precisely, notice
that the first 5 component fields of $\Phi + \Phi ^{\dagger}$
in Eq.~\ref{eq:compphi} are completely unconstrained.
This means that without loss of generality we can decompose
any vector superfield as follows:
\be
V\sx = V_{WZ} + \Phi + \Phi ^{\dagger} \quad ,
\ee
where $V_{WZ}$ only has 4 component fields instead of 9:
\be
V_{WZ} = -\th\sigma ^m\barth v_m +
i(\th\th )\barth\barlam - i(\barth\barth )\th\lambda
+ \half (\th\th )(\barth\barth )D \quad ,
\label{eq:vwzdef}
\ee
where, to conform with Wess and Bagger, I have changed
notation slightly:
\begin{eqnarray*}
v_m &\rightarrow & -v_m \quad ,\\
\barlam &\rightarrow & i\barlam \quad ,\\
d &\rightarrow & \half D \quad .
\end{eqnarray*}
$V_{WZ}$ is known as the {\bf Wess-Zumino gauge-fixed superfield}.

This decomposition is unambiguous {\bf except} for the
remaining freedom to shift part of $v_m$ into the corresponding
component of $\Phi + \Phi ^{\dagger}$, i.e.
$v_m \rightarrow v_m$$-$$i\partial _m(A$$-$$A^*)$.
Thus fixing Wess-Zumino gauge {\bf does not} fix the
abelian gauge freedom.

\subsection{The supersymmetric field strength}

Note that the supersymmetry transformations do not respect
the Wess-Zumino gauge-fixing decomposition. This is somewhat
disappointing since it means that a superfield formulation
in terms of $V\sx$ necessarily carries around a number of
superfluous fields. We can however define a different superfield
which has the property that it only contains the Wess-Zumino
gauge-fixed component fields $v_m(x)$, $\lambda (x)$, and $D(x)$.

We define left and right-handed {\bf spinor superfields}
$\W$, $\Wdot = (\W )^{\dagger}$:
\bea
\W &=& -\quarter (\barD\barD )\D V\sx \quad ,\nonumber\\
\Wdot &=& -\quarter (DD) \Ddot V\sx \quad .
\label{eq:simplewdef}
\eea
An equivalent definition, which we will need when we
go from the abelian to the nonabelian case, is:
\bea
\W &=& -{1\over 8}(\barD\barD )\e{-2V}\D \e{2V}
\quad ,\nonumber\\
\Wdot &=& {1\over 8}(DD)\e{2V}\Ddot \e{-2V} \quad .
\label{eq:nonawdef}
\eea

$\W$ is a chiral superfield:
\be
\Ddot W_{\beta} = -\quarter \Ddot (\barD _{\gammadot}
\barD ^{\gammadot})D_{\beta}V\sx = 0\quad ,
\ee
where we have used the fact that since the $\barD$'s
anticommute and have only 2 components, $(\barD )^3 = 0$.

$\Wdot$ is an antichiral superfield.
$\W$ is not a general chiral spinor superfield, because
$\W$ and $\Wdot$ are related by an
additional covariant constraint:
\be
\Ddot \barW ^{\alphadot} = D^{\alpha}\W \quad .
\label{eq:addcovw}
\ee
This constraint follows trivially from Eq.~\ref{eq:simplewdef}:
\bea
\Ddot \barW ^{\alphadot} =& \epsilon ^{\alphadot\betadot}
\Ddot \barW _{\betadot} &= -\quarter\epsilon ^{\alphadot\betadot}
\Ddot (DD)\barD _{\betadot}V \nonumber\\
=& -\quarter (\barD\barD )(DD)V &= -\quarter D^{\alpha}
(\barD\barD)\D V \\
=& D^{\alpha}\W &\quad .\nonumber
\eea
$\W$ and $\Wdot$ are both invariant under the transformation
Eq.~\ref{eq:abelgt}. Let us prove this for $\W$:
\bea
\W &\rightarrow & -\quarter (\barD\barD )\D
(V + \Phi + \Phi ^{\dagger})
\quad ,\nonumber\\
&=& \W -\quarter (\barD\barD )\D\Phi\quad ,\quad
({\rm since\ }\D\Phi ^{\dagger} = 0)\nonumber\\
&=& \W +\quarter \barD ^{\betadot} \{ \barD _{\betadot},
\D \}\Phi\quad ,\quad ({\rm since\ }\barD _{\betadot}\Phi = 0)
\label{eq:howwt}\\
&=& \W \quad ,\nonumber
\eea
where in the last step we have used:
\bea
\{ \barD _{\betadot},\D \} &=& -2\sigma ^m_{\alpha\betadot}P_m
\quad ,\nonumber\\
{[}\, \barD ^{\betadot},P_m \,{]} &=& 0 \quad .
\label{eq:howwtid}
\eea

Since $\W$ and $\Wdot$ are both invariant under Eq.~\ref{eq:abelgt},
there is no loss of generality in computing their components
in Wess-Zumino gauge, i.e. write
\bea
\W &=& -\quarter (\barD\barD )\D V_{WZ}\sx \quad ,\nonumber\\
\Wdot &=& -\quarter (DD) \Ddot V_{WZ}\sx \quad .
\label{eq:otherwdef}
\eea
Since $\W = \W\sy$ we write
\be
V_{WZ}\sx = V_{WZ}(y - i\th\sigma\barth, \th ,\barth )
\ee
and expand $\W$ in component fields which are functions of $y$:
\bea
\W &=& -i\lambda _{\alpha}(y) + \th _{\alpha}D(y)
-{i\over 2}(\sigma ^m\barsigma ^n\th )_{\alpha}
(\partial _mv_n - \partial _nv_m)(y)\nonumber\\
&&\quad+(\th\th )\sigma ^m_{\alpha\betadot}\partial _m
\barlam ^{\betadot}(y) \quad ,
\label{eq:abelwcom}\\
\Wdot &=& i\barlam _{\alphadot}(y^{\dagger}) +
\barth _{\alphadot}D(y^{\dagger})
+{i\over 2}(\barsigma ^m\sigma ^n\barth )_{\alphadot}
(\partial _mv_n - \partial _nv_m)(y^{\dagger})\nonumber\\
&&\quad-(\barth\barth )
\hbox{$\barsigma ^m_{\alphadot}$}^{\beta}\partial _m
\lambda _{\beta}(y^{\dagger}) \quad .\nonumber
\eea
So indeed $\W$, $\Wdot$ contain only the component fields
\[
\lambda ,\;D,\; f_{mn}\equiv\partial _mv_n -\partial _nv_m \quad .
\]
This is an irreducible off-shell multiplet known as
the {\bf curl multiplet} or {\bf field strength multiplet};
it has $4$$+$$1$$+$$3 = 8$ real components.

\subsubsection{Nonabelian generalization}

We can exponentiate the infinitesimal abelian transformation
Eq.~\ref{eq:abelgt} to obtain the finite transformation
\be
\e{V} \rightarrow \e{-i\Lambda ^{\dagger}}\e{V}
\e{i\Lambda} \quad ,
\label{eq:nonabt}
\ee
where, to conform with the standard notation of
Ferrara and Zumino, \cite{ferrara}
we now denote the chiral superfields
of Eq.~\ref{eq:abelgt} by:
\begin{eqnarray*}
\Phi &\rightarrow& i\Lambda \quad ,\\
\Phi ^{\dagger} &\rightarrow& -i\Lambda ^{\dagger} \quad .
\end{eqnarray*}
To obtain the nonabelian generalization we write
\bea
V &\rightarrow& T_{ij}^aV_a \quad ,\nonumber\\
\Lambda &\rightarrow& T_{ij}^a\Lambda _a \quad ;\\
{[}\, T^a,T^b \,{]} &=& if^{abc}T^c \quad ,\nonumber\\
\tr T^aT^b &=& \delta ^{ab}\quad ,\nonumber
\eea
where the $T^a_{ij}$ are the hermitian generators of some
Lie algebra.
The form of the nonabelian transformation is then the same as
Eq.~\ref{eq:nonabt}.

To find the infinitesimal nonabelian transformation, we
can apply the Baker-Campbell-Hausdorff formula to
Eq.~\ref{eq:nonabt}. One can show that, to first order
in $\Lambda$, Eq.~\ref{eq:nonabt} reduces to: \cite{rocek}
\be
\delta V = iL_{V/2}(\Lambda + \Lambda ^{\dagger})
+iL_{V/2}{\rm coth}L_{V/2}
(\Lambda - \Lambda ^{\dagger}) \quad ,
\label{eq:fullV}
\ee
where the operation $L_XY$ denotes the Lie derivative:
\bea
L_XY &=& [X,Y] \quad ,\nonumber\\
(L_X)^2Y &=& [X,[X,Y]] \quad ,\\
&{\rm etc.}& \quad .\nonumber
\eea
Eq.~\ref{eq:fullV} is meant to be evaluated by its
Taylor series expansion, using
\be
x\,{\rm coth}x = 1 + {x^2\over 3} - {x^4\over 45} + \ldots \quad .
\ee

This becomes much more illuminating if we fix the nonabelian
equivalent of Wess-Zumino gauge. Unlike the abelian case,
the relationship between the component fields of $V\sx$
and $\Lambda\sy$ in the Wess-Zumino gauge fixing is nonlinear,
due to the complicated form of Eq.~\ref{eq:fullV}. However the
end result is the same: $V_{WZ}\sx$ is as given in
Eq.~\ref{eq:vwzdef}. Furthermore, as in the abelian case,
the Wess-Zumino decomposition does not fix the freedom
to perform gauge transformations parametrized by the scalar
component of $\Phi$$-$$\Phi ^{\dagger} \equiv i(\Lambda$$+$$
\Lambda ^{\dagger})$.

Consider then the transformation Eq.~\ref{eq:fullV} with
$V$ replaced by $V_{WZ}$, and with only the scalar component
of $\Lambda + \Lambda ^{\dagger}$ nonvanishing (which also
implies that only the $\th\sigma\barth$ component of
$\Lambda - \Lambda ^{\dagger}$ is nonvanishing). Clearly
only the first term in the Taylor series expansion of
the hyperbolic cotangent remains, since the next higher
order term gives something proportional to $\th ^3 \barth ^3$.
Thus having fixed Wess-Zumino gauge the infinitesimal
nonabelian gauge transformation is just
\be
\delta V = i(\Lambda - \Lambda ^{\dagger} )
-{i\over 2}[(\Lambda + \Lambda ^{\dagger}),V] \quad.
\label{eq:inagt}
\ee
This implies the usual nonabelian gauge transformations
for the component fields $v_m(x)$, $\lambda (x)$,
and $D(x)$ ($v_m(x)$ is the nonabelian gauge field while
$\lambda (x)$ and $D(x)$ are matter fields in the adjoint
representation).

$\W$ and $\Wdot$ are given by Eq.~\ref{eq:nonawdef} in the
nonabelian case. Let us compute how $\W$ and $\Wdot$ transform
under Eq.~\ref{eq:nonabt}. First notice that, under
the transformation Eq.~\ref{eq:nonabt}:
\be
\e{-2V}\D \e{2V} \rightarrow \e{-i2\Lambda}
\e{-2V}\left( \D\e{2V} \right) \e{i2\Lambda}
+ \e{-i2\Lambda}\D\e{i2\Lambda} \quad ,
\ee
which follows from the fact that $\D\Lambda ^{\dagger} = 0$.
Thus, using also the fact that $\Ddot$ commutes with
$\Lambda$, we see that
\be
\W \rightarrow \e{-i2\Lambda}\W\e{i2\Lambda} -
{1\over 8}\e{-i2\Lambda}(\barD\barD )\D\e{i2\Lambda}
\quad .
\ee
Furthermore the second term vanishes, just as in
Eq.~\ref{eq:howwt}, using the identities Eq.~\ref{eq:howwtid}.
So our final result is that $\W$ and $\Wdot$ transform
covariantly in the nonabelian case:
\bea
\W \rightarrow \e{-i2\Lambda}\W\e{i2\Lambda} \quad ,
\nonumber\\
\Wdot \rightarrow
\e{-i2\Lambda ^{\dagger}}\Wdot\e{i2\Lambda ^{\dagger}} \quad .
\eea

Let us be more explicit in the nonabelian case about
the derivation of the component expansion for $\W$.
There is no loss of generality in computing this in Wess-Zumino gauge.
{}From the definition Eq.~\ref{eq:nonawdef} we have the
explicit expression:
\bea
\W &=&-{1\over 8}(\barD\barD )\e{-2V_{WZ}}\D \e{2V_{WZ}}
\quad ,\\
&=& -\quarter (\barD\barD )\D V_{WZ}
+ \half (\barD\barD )V_{WZ}\D V_{WZ}
- \quarter (\barD\barD)\D V_{WZ}^2\nonumber
\eea
where we have used our knowledge (see Eq.~\ref{eq:vwzdef}) of the
component expansion for $V_{WZ}(y-i\th\sigma\barth ,\th ,\barth )$:
\bea
V_{WZ}(y-i\th\sigma\barth ,\th ,\barth ) &=& -\th\sigma ^m\barth v_m(y) +
i(\th\th )\barth\barlam (y) - i(\barth\barth )\th\lambda (y) \nonumber\\
&&+ \half (\th\th )(\barth\barth )\left( D(y) + i\partial ^mv_m(y)
\right) \quad .
\label{eq:newvwzdef}
\eea
Using the form Eq.~\ref{eq:ycovderdef} for $\D$ acting on
functions of $(y,\th ,\barth )$, we have:
\bea
\lefteqn{\D V_{WZ} = -\sigma ^m_{\alpha\betadot}\barth ^{\betadot}
v_m(y) + 2i\th _{\alpha}\barth\barlam (y)
-i(\barth\barth )\lambda _{\alpha}(y)
}\nonumber\\
&&+\th _{\alpha}(\barth\barth )\left( D(y) + i\partial ^mv_m(y)
\right)
-i(\barth\barth )(\sigma ^m\barsigma ^n\th )_{\alpha}
\partial _mv_n(y)\\
&&+(\th\th )(\barth\barth )\sigma ^m_{\alpha\betadot}\partial _m
\barlam ^{\betadot} \quad .\nonumber
\eea
A little more straightforward computation gives:
\be
\D V_{WZ}^2 = \th _{\alpha}(\barth\barth )v^mv_m
\quad ,
\ee
as well as:
\bea
V_{WZ}\D V_{WZ} &=& \half\th _{\alpha}(\barth\barth )v^mv_m(y)
+\quarter\sigma ^n_{\alpha\betadot}\barsigma ^{m\betadot\gamma}
\th _{\gamma}(\barth\barth )[v_m,v_n] \nonumber\\
&&-\half i(\th\th )(\barth\barth )\sigma^m_{\alpha\betadot}
[v_m,\barlam ^{\betadot}]
\quad .
\eea
Putting it all together, we have:
\be
\W = -i\lambda _{\alpha}(y) + \th _{\alpha} D(y)
-\sigma _{\alpha}^{mn\beta}\th _{\beta}F_{mn}(y)
+(\th\th )\sigma ^m_{\alpha\betadot}\grad{m}\barlam ^{\betadot}(y)
\quad ,
\label{eq:wdown}
\ee
where
\bea
F_{mn} &=& \partial _mv_n - \partial _nv_m +i[v_m,v_n]
\quad ,\nonumber\\
\grad{m}\barlam ^{\betadot} &=& \partial _m\barlam ^{\betadot}
+i[v_m,\barlam ^{\betadot}]
\quad ;
\eea
$F_{mn}$ is the Yang-Mills field strength, while $\grad{m}$
is the Yang-Mills gauge covariant derivative.

We also need
\be
W^{\alpha} = -{1\over 8}(\barD\barD )\e{-2V}D^{\alpha} \e{2V}\quad ;
\ee
raising the index on Eq.~\ref{eq:wdown} and Fierzing, we get:
\be
W^{\alpha} = -i\lambda ^{\alpha}(y) + \th ^{\alpha} D(y)
+\th ^{\beta}\sigma _{\beta}^{mn\alpha}F_{mn}(y)
-(\th\th )\barsigma ^{m\betadot\alpha}\grad{m}\barlam _{\betadot}(y)
\quad .
\label{eq:wup}
\ee

\subsection{N=1 linear multiplet}

In the previous subsection we obtained the field strength
multiplet by starting with the chiral spinor
superfield $\W$, and imposing the additional
covariant constraint Eq.~\ref{eq:addcovw}.
Let us again start with a chiral spinor superfield
$\Phi _{\alpha}$, $\barPhi _{\alphadot} =
(\Phi _{\alpha})^{\dagger}$, and construct a new
superfield $L\sx$ as follows:
\be
L\sx = i\left( D^{\alpha}\Phi _{\alpha} +
\barD _{\alphadot} \barPhi ^{\alphadot} \right)
\quad .
\ee
The superfield $L\sx$ is real, since
\be
\left( \Ddot \barPhi ^{\alphadot} \right)^{\dagger}
= \D\Phi ^{\alpha} = -D^{\alpha}\Phi _{\alpha}
\quad ;
\ee
so $L\sx$ is a vector superfield which satisfies
two additional covariant constraints:
\be
(DD)L = (\barD\barD )L = 0 \quad .
\ee
These constraints follow trivially from the fact that
$\Phi$ is chiral, and $(D)^3  = (\barD )^3 = 0$.

The component fields of $L\sx$ comprise the
{\bf linear multiplet}. These are a real scalar $C(x)$,
a complex left-handed Weyl spinor $\chi _{\alpha}$,
and a real divergenceless vector field $A_m$,
$\partial ^mA_m = 0$. Thus the linear multiplet has
$1$$+$$4$$+3 = 8$ real components.

\section{N=1 Globally Supersymmetric Actions}

Recall from the previous section that both the $F$
component of a chiral superfield and the $D$ component
of a vector superfield transform by a total derivative
under an N=1 supersymmetry transformation. Thus we
immediately deduce two classes of N=1 globally
supersymmetric actions:
\be
\int d^4x\; \left[ \int d^2\th\, \Phi\sy +
\int d^2\barth\, \Phi ^{\dagger}(y^{\dagger},\barth )
\right]
\ee
is an invariant real action for {\bf any} chiral
superfield $\Phi\sy$, while
\be
\dff V\sx
\ee
is an invariant real action for {\bf any} vector
superfield $V\sx$.

\subsection{Chiral superfield actions}

The Wess-Zumino model \cite{wz} is the simplest (sensible)
N=1 SUSY model in four dimensions. The action is
\be
\dff \Phi ^{\dagger}\Phi -
\df \left[ \dt (\half m\Phi ^2 + \third g\Phi ^3 ) +{\rm h.c.}
\right] \quad ,
\ee
where $\Phi$ is a chiral superfield.

Let us work out the part of this action containing
bosonic component fields. The bosonic components of
$\Phi$ and $\Phi ^{\dagger}$ are:
\bea
\Phi\sy &=& A(x) + \th\th F(x) + i\th\sigma ^m\barth
\partial _mA(x)\nonumber\\
&&-\quarter (\th\th )(\barth\barth )\BBox A(x)
\quad ,\nonumber\\
\Phi ^{\dagger}(y^{\dagger},\barth ) &=& A^*(x) +
\barth\barth F^*(x) - i\th\sigma ^m\barth\partial _mA^*(x)
\\
&&-\quarter (\th\th )(\barth\barth )\BBox A^*(x) \quad .\nonumber
\eea
Thus:
\be
\Phi ^{\dagger}\Phi \Bigl|_{\th\th\barth\barth} =
-\quarter\BBox A^*A -\quarter A^*\BBox A + F^*F
+\half\partial ^mA^*\partial _mA \quad ,
\label{eq:ppdag}
\ee
where to obtain the last term we have used the
Fierz identity Eq.~\ref{eq:goodfierz}.

We also have
\be
\left[ \half m\Phi ^2 + \third g\Phi ^3 \right]_{\th\th}
= mAF + gA^2F \quad ,
\ee
so the part of the Wess-Zumino action containing only
bosonic fields is:
\be
\df \left[ \partial ^mA^*\partial _mA + F^*F
-( mAF + gA^2F + {\rm h.c.}) \right]
\ee

We immediately notice that this action contains no
derivatives acting on $F(x)$, i.e. $F(x)$ is an
{\bf auxiliary field} which can be eliminated by
solving its equations of motion:
\bea
{\delta\lag\over\delta F} &=&
F^* - mA - gA^2 = 0 \quad ,\nonumber\\
{\delta\lag\over\delta F^*} &=&
F - mA^* - g(A^*)^2 = 0 \quad .
\eea
This means we can write the bosonic part of the Wess-Zumino
action as just
\be
\df \left[ \partial ^mA^*\partial _mA -
V(A,A^*) \right] \quad ,
\ee
where the scalar potential $V(A,A^*)$ is given by:
\be
V(A,A^*) = \vert F \vert ^2 =
[mA^* + g(A^*)^2][mA + gA^2] \quad .
\ee

More generally we could write
\be
\dff \Phi ^{\dagger}\Phi - \df \left[ \dt
W(\Phi ) + {\rm h.c.} \right] \quad ,
\label{eq:supact}
\ee
where the {\bf superpotential} $W(\Phi )$ is a
{\bf holomorphic function} of $\Phi$, i.e. a functional
only of $\Phi$, not $\Phi ^{\dagger}$.
In this more general case the scalar potential is
\be
V_F(A,A^*) = \vert F \vert ^2 =
\Bigl| {\delta W\over\delta\Phi} \Bigr|^2_{\Phi = A}
\quad .
\ee
Note that the scalar potential is obviously positive definite.

Since a cubic superpotential leads to a quartic scalar potential,
we also see that the Wess-Zumino model is the most general unitary,
{\bf renormalizable} four-dimensional SUSY action for a
single chiral superfield.

An even more general construction than Eq.~\ref{eq:supact} is
\be
\lag = \dtf K(\Phi ^i,\hbox{$\Phi ^j$}^{\dagger})
- \left[ \dt W(\Phi ^i ) + {\rm h.c.} \right] \quad ,
\label{eq:genkahact}
\ee
where $K(\Phi ^i,\hbox{$\Phi ^j$}^{\dagger})$ is
called the {\bf \Kahler\ potential}, and we now have an
arbitrary number of chiral superfields $\Phi ^i$.
The \Kahler\ potential
is a vector superfield; unlike the superpotential it is
obviously not a holomorphic function of the $\Phi ^i$.

{}From the component expansion Eq.~\ref{eq:fullchiexp}
it is clear that the \Kahler\ potential produces
kinetic terms with no more than two spacetime derivatives.
If we replace some of the $\Phi ^i$ by covariant
derivatives of superfields, we will either obtain a higher
derivative theory, or a theory which can be collapsed back
to the form Eq.\ref{eq:genkahact}. Thus if we exclude
higher derivative theories Eq.\ref{eq:genkahact} is
the most general action for (not necessarily renormalizable)
N=1 SUSY models constructed from chiral superfields.

\subsection{N=1 supersymmetric nonlinear sigma models}

Bosonic nonlinear sigma models in $D$-dimensional
spacetime have an action of the form:
\be
\half\int d^Dx\; g_{ij}(A)\; \partial ^mA^i(x) \partial _mA^j(x) \quad ,
\ee
where the $A^i(x)$ are real scalar fields. The functional
$g_{ij}(A)$ can be thought of as the {\bf metric} of a
{\bf target space}
Riemannian manifold with line element
\be
ds^2 = g_{ij}\, dA^idA^j \quad .
\label{eq:linel}
\ee
Nonlinear sigma models are not in general renormalizable,
except in the case $D=2$ with $g_{ij}$ the metric of a
symmetric space. \cite{friedan}

The general chiral superfield action Eq.\ref{eq:genkahact}
defines the supersymmetrized version of 4-dimensional
nonlinear sigma models. \cite{zumsig}
To see this, note that Eq.~\ref{eq:ppdag}
implies that the kinetic term for the complex scalar
components $A^i(x)$ is
\be
g_{ij*}\, \partial _m A^i(x) \partial ^mA^{*j} \quad,
\ee
where:
\vskip .1in
\be
g_{ij*} \equiv {\delta ^2K(A^i,A^{*j})\over
\delta A^i\delta A^{*j} } \quad .
\label{eq:kast}
\ee
\vskip .1in
Since the \Kahler\ potential is real, the target space metric
$g_{ij*}$ is hermitian. To obtain a correct sign kinetic term
for every nonauxiliary scalar field, we must also require
that $g_{ij*}$ is positive definite and nonsingular;
this implies (mild) restrictions on the choice of
the \Kahler\ potential.

A complex Riemannian manifold possessing a positive definite
nonsingular hermitian metric which can be written (locally)
as the second derivative of a
scalar function is called a {\bf \Kahler\ manifold}.
Thus Eq.\ref{eq:genkahact} defines supersymmetric
generalized nonlinear sigma models whose target spaces are
\Kahler\ manifolds.

This is a rather powerful observation, since it implies
that models with horrendously complicated component field
Lagrangians can be characterized by the algebraic geometry
of the target space. As an example, we will discuss the
possible {\bf holonomy groups} of sigma model target
spaces.

Consider the parallel transport of a vector around a
contractible closed loop using the Riemannian connection
in a $D$-dimensional Riemannian space. The transported
vector is related to the original vector by some $SO(D)$
rotation. The $SO(D)$ matrices obtained this way form a
group, the local holonomy group of the manifold. Obviously
the holonomy group is either $SO(D)$ itself or a subgroup
of it. Four important examples are given below
(we use the convention that $Sp(2D)$ is the
symplectic group of rank $D$):

\goodbreak
\vskip .2in
\begin{flushleft}
{\bf Manifold\hfill Maximum Holonomy Group}\\
\ \\
General Riemannian space with real dimension $D$:\leaderfill $SO(D)$\\
\ \\
\Kahler\ manifold with complex dimension $D$,\\
real dimension $2D$:\leaderfill $U(D)$\\
\ \\
Hyper\Kahler\ manifold with real dimension $4D$:
\leaderfill $Sp(2D)$\\
\ \\
Quaternionic manifold with real dimension $4D$:
\leaderfill $Sp(2D)\times Sp(2)$\\
\ \ \
\end{flushleft}
\vskip .2in

Note that the \Kahler\ structure Eq.~\ref{eq:kast}
(and thus also the action) is invariant under a
{\bf \Kahler\ transformation}:
\be
K(A^i,A^{*j}) \rightarrow
K(A^i,A^{*j}) + \Lambda (A^i) + \Lambda ^{\dagger}(A^{*j})
\quad .
\ee
It is also clear that both the
\Kahler\ structure Eq.~\ref{eq:kast} and the Riemannian structure
Eq.~\ref{eq:linel} are preserved by arbitrary {\bf holomorphic}
transformations of the target space coordinates $A^i$.

\subsection{N=1 supersymmetric Yang-Mills theory}

We recall that $\W$ is a chiral spinor superfield
and that a gauge transformation on the vector component
of $\W$ is induced by the superfield transformation
\be
\W \rightarrow \e{-i2\Lambda}\W\e{+i2\Lambda} \quad .
\ee
It follows that a gauge invariant supersymmetric action is
\bea
&&\half \df\dt \tr W^{\alpha}W_{\alpha} \\
&& = \df \tr \left[ -\quarter F_{mn}F^{mn}
- {i\over 4}F_{mn}\tilde{F}^{mn} -
i\lambda\sigma ^m\grad{m}\barlam + \half D^2 \right]
\quad ,\nonumber
\eea
where we have used the explicit component expansions
Eqs.~\ref{eq:wdown},\ref{eq:wup}. The dual field
strength is defined as:
\be
\tilde{F}^{mn} \equiv \half\epsilon ^{mnpq}F_{pq}
\quad .
\ee
This action is not real and lacks any dependence upon
the Yang-Mills gauge coupling $g$. The duality-friendly
way to remedy these deficiencies is by introducing a complex
gauge coupling $\tau$:
\be
\tau = {\tym\over 2\pi} + {4\pi i\over g^2} \quad ,
\ee
where $\tym$ is the Yang-Mills theta parameter. The N=1
Yang-Mills action we want is then
\bea
&&{1\over 8\pi}{\rm Im}\,\left[ \tau \df\dt \tr W^{\alpha}W_{\alpha}
\right] \\
&& = {1\over g^2}\df \tr \left[ -\quarter F_{mn}F^{mn}
-i\lambda\sigma ^m\grad{m}\barlam + \half D^2 \right]
\\
&&\qquad\quad - {\tym\over 32\pi ^2}\df \tr F_{mn}\tilde{F}^{mn}
\quad .\nonumber
\eea
The minus sign in front of the $\tym$ term is correct given
the minus sign convention of Eq.~\ref{eq:gengroupel}.

Under a gauge transformation, chiral superfields
$\Phi$ in the adjoint representation transform as:
\bea
\Phi &\rightarrow& \e{-i2\Lambda}\Phi
\quad ,\nonumber\\
\Phi ^{\dagger} &\rightarrow& \Phi ^{\dagger}\e{i2\Lambda ^{\dagger}}
\quad .
\eea
Thus $\tr \Phi ^{\dagger}\Phi$ is not gauge invariant.
However from Eq.~\ref{eq:inagt} we see that the following
is a gauge invariant kinetic term for chiral superfields:
\be
\tr \Phi ^{\dagger} \e{2V} \Phi \quad .
\label{eq:gkt}
\ee
In fact this is gauge invariant for $\Phi$ in an
arbitrary representation $R$, not just the adjoint.
In this case $\Lambda$$=$$t^a_{ij}\Lambda _a$, where
the $t^a_{ij}$ are matrices in the representation $R$.
Thus Eq.~\ref{eq:gkt} is still gauge invariant
provided that all the tensor products contain the singlet.
This is indeed true because the tensor product of $R$ with its
conjugate $\overline{R}$ contains both the singlet and
adjoint representations, while every term in the series
expansion of exp($2V$) also contains either the singlet
or the adjoint (or both).

Thus, supposing I have chiral superfields $\Phi ^i$
transforming in representations $R_i$, the gauged
version of the Wess-Zumino action is:
\bea
\lefteqn{ \tr \dff \hbox{$\Phi ^i$}^{\dagger}\e{2V}\Phi ^i }
\nonumber\\
&&-\tr\df \left[ \dt (\half m _{ij}\Phi ^i\Phi ^j
+ \third g_{ijk}\Phi ^i\Phi ^j\Phi ^k ) +{\rm h.c.}
\right] \quad .
\eea
Note that by gauge invariance $m_{ij}$ can only
be nonvanishing if
\be
R_i = {\overline{R}}_j \quad .
\ee
Similarly, $g_{ijk}$ can only be nonvanishing if
$R_i$$\times$$R_j$$\times$$R_k$ contains the singlet.

The gauged kinetic term Eq.~\ref{eq:gkt} contains
a D-term $A^*DA$. The only other dependence on
the auxiliary field $D$ is the term $D^2/2g^2$ in
the Yang-Mills action. Thus when we eliminate this
auxiliary field by its equation of motion we find
\bea
D_a &=& -g^2A_bA^*_c\,\tr (T^aT^bT^c) \nonumber\\
&=& -\half ig^2f^{abc}A_bA^*_c \quad ,
\eea
where the second line follows from the fact that
the adjoint representation is always anomaly-free.
Thus:
\be
D = T^aD_a = \half g^2[A,A^*] \quad .
\ee
This implies that in the coupled Yang-Mills-Wess-Zumino
model there is a new contribution to the scalar potential:
\be
V_D = {1\over 2g^2}D^2 =
{g^2\over 8}\left( [A^i,\hbox{$A^i$}^*] \right)^2\quad .
\label{eq:vd}
\ee
So altogether the complete scalar potential is the sum
of positive definite F and D-term contributions:
\be
V(A^i,\hbox{$A^i$}^*) = V_F + V_D = \vert F \vert ^2 +
{1\over 2g^2}D^2
\ee

If we forget about renormalizablity we can write
a very general N=1 action by gauging Eq.~\ref{eq:genkahact}:
\bea
\lefteqn{\lag = \dtf K(\Phi ^i,\hbox{$\Phi ^j$}^{\dagger}\e{2V})
- \left[ \dt W(\Phi ^i ) + {\rm h.c.} \right] }\nonumber\\
& & +{1\over 8\pi}{\rm Im}\,\left[ \tau \dft \tr f(\Phi ^i)
W^{\alpha}W_{\alpha}
\right]
\quad ,
\label{eq:mostgenact}
\eea
where $f(\Phi ^i)$ is a new holomorphic function called
the {\bf gauge kinetic function}. Note that every term
in $f(\Phi ^i)$ must transform like a representation
which is contained in the tensor product of two adjoints.

\section{N=2 Globally Supersymmetric Actions}

\subsection{N=2 superspace}

There are several different ways to extend our
treatment of N=1 rigid superspace to the case of
N=2 rigid superspace. \cite{supertwo} Some methods,
e.g. {\bf harmonic superspace}, build in the $SU(2)$
automorphism symmetry of the N=2 generators $Q_{\alpha}^1$,
$Q_{\alpha}^2$.

We will make do with the most naive extension of
N=1 to N=2 superspace parameterizations:
\bea
\tha ,\;\bartha &\rightarrow &
\tha ,\; \bartha ,\; \that ,\; \barthat
\nonumber\\
\D &\rightarrow & \D ,\; \Dta
\nonumber\\
\Ddot &\rightarrow & \Ddot ,\; \Ddott
\\
\dt &\rightarrow & \dtt
\nonumber
\eea

If we want to restore the $SU(2)$ global R symmetry, we
should think of $(\tha ,\that )$ etc. as $SU(2)$ doublets.

\subsection{N=2 chiral superfields}

An N=2 chiral superfield $\Psi\sxt$ is defined as an
N=2 scalar superfield which is a singlet
under the global $SU(2)$ and which satisfies the covariant constraints
\bea
\Ddot \Psi\sxt &=& 0
\quad ,\nonumber\\
\Ddott \Psi\sxt &=& 0
\quad .
\eea
It is convenient to introduce new bosonic coordinates
\be
\yt ^m = x^m + i\th\sigma ^m\barth + i \tht\sigma ^m\bartht
\quad ,
\ee
which obviously satisfy
\be
\Ddot \yt ^m = \Ddott \yt ^m = 0 \quad .
\ee
If we expand an N=2 chiral superfield in powers of $\tht$,
the components are N=1 chiral superfields. Thus:
\be
\Psi = \Phi\syt + i\sqrt{2}\that\W\syt + \tht\tht G\syt
\quad,
\ee
where $\Phi (y$$+$$i \tht\sigma ^m\bartht, \th )$
and $G(y$$+$$i \tht\sigma ^m\bartht, \th )$ are N=1
chiral superfields (note the effective ``$y$'' coordinate
is shifted by $i\tht\sigma ^m\bartht$),
and $\W (y$$+$$i \tht\sigma ^m\bartht, \th )$
is an N=1 chiral spinor superfield.

Since $\Psi$ is an $SU(2)$ singlet, while
$(\tha ,\that )$ is an $SU(2)$ doublet, it follows
that the fermionic components $\psi$ of $\Phi$
and $\lambda$ of $\W$ also form an $SU(2)$ doublet.
On the other hand the bosonic component fields $A$ of
$\Phi$ and $v_m$ of $\W$ are $SU(2)$ singlets.

\subsection{N=2 supersymmetric Yang-Mills theory}

Suppose we write
\be
\Psi\syt \rightarrow T^a_{ij}\Psi _a\syt \quad .
\ee
Then, since
\be
\Psi ^2 \Bigl|_{\th\th\tht\tht} =
W^{\alpha}W_{\alpha}\Bigl|_{\th\th} +
2G\Phi\Bigl|_{\th\th}
\label{eq:myguy}
\ee
the obvious form for N=2 Yang-Mills theory is:
\be
{1\over 4\pi}{\rm Im}\,\left[ \tau \df\!\!\dtt \tr \,\half\Psi ^2\,
\right] \quad .
\label{eq:ntwoact}
\ee
This clearly describes an N=1 Yang-Mills theory coupled
to chiral superfields in the adjoint representation.
Unfortunately something is wrong, since the second term
in Eq.~\ref{eq:myguy} is not a sensible Lagrangian
for chiral superfields. Clearly what we want is to be
able to regard $G\syt$ as an auxiliary superfield,
and thus eliminate it in favor of $\Phi$ and $V$, reproducing
(at least) the N=1 gauge invariant kinetic term
Eq.~\ref{eq:gkt}.

Thus, while Eq.~\ref{eq:ntwoact} is the correct action
for N=2 Yang-Mills theory, we must impose additional
covariant constraints on the N=2 chiral superfield
$\Psi$. The correct constraints turn out to be:
\be
(D^{a\alpha}D_{\alpha}^b)\Psi =
(\barD^a_{\alphadot}\barD^{b\alphadot})\Psi ^{\dagger}
\quad ,
\label{eq:twocc}
\ee
where $a,b$ are global $SU(2)$ indices:
\bea
D^a &=& (D,\Dt ) \quad ,\nonumber\\
\barD ^a &=& (\barD ,\barDt ) \quad .
\eea

Rather than solve these constraints directly,
it is much easier to simply assume that $G$ can
be eliminated in favor of $\Phi$ and $V$, then
deduce the correct expression from the requirement
of gauge invariance (i.e. gauge invariance in the N=1 sense).
Roughly speaking, we need something like
\be
G\syt \sim \Phi ^{\dagger}\e{2V} \quad .
\ee
However, while the right-hand side transforms correctly
under gauge transformations, it is clearly {\bf not}
an N=1 chiral superfield. So consider instead the
more sophisticated expression:
\be
G\syt = \int d^2\barth\; \Phi ^{\dagger}
(\yt - i\th\sigma\barth ,\barth )
\e{2V(\yt - i\th\sigma\barth ,\th ,\barth )}
\quad ,
\label{eq:gjunk}
\ee
where the integral is meant to be performed for
{\bf fixed} ${\yt}$.

The result of the integral is obviously a function
only of $\yt$ and $\th$, so $G\syt$ thus defined is
an N=1 chiral superfield, as required.
Under the N=1 superfield transformation which induces
a gauge transformation, the integrand of Eq.~\ref{eq:gjunk}
transforms as:
\bea
\Phi ^{\dagger}\e{2V} &\rightarrow & \Phi ^{\dagger}
\e{2V}\e{i2\Lambda (\yt - i\th\sigma\barth + i\th\sigma\barth ,\th )}
\nonumber\\
&=& \Phi ^{\dagger}\e{2V}\e{i2\Lambda (\yt ,\th )}
\quad ,
\eea
so we can pull the exp$(i2\Lambda\syt$ factor out of the integral.
Thus
\be
G\syt \rightarrow G\syt\e{i2\Lambda\syt}
\quad ,
\ee
as required for gauge invariance.

The overall coefficient of 1 in Eq.~\ref{eq:gjunk}
is fixed by the global $SU(2)$ symmetry. As we noted above
the fermionic components $\psi$ of $\Phi$
and $\lambda$ of $\W$ form an $SU(2)$ doublet.
Thus the relative coefficient of the kinetic terms
for $\psi$ in $G\Phi$ and $\lambda$ in $W^{\alpha}\W$
must be equal.

The resulting N=2 Yang-Mills theory is thus equivalent
to N=1 Yang-Mills coupled to matter fields in the adjoint representation.
There is no superpotential, but there is a scalar potential
coming from the D-term. The nonauxiliary fields form an
off-shell N=2 vector multiplet: $v_m$, $A$, and the global
$SU(2)$ doublet $(\psi ,\lambda )$. On-shell this multiplet
gives $4$$+$$4 = 8$ real field components, which of course
agrees with the counting for the massless N=2 vector multiplet
of single particle states.

\subsection{The N=2 prepotential}

If we forget about renormalizability, we can write a much more
general action for N=2 chiral superfields satisfying the
covariant constraint Eq.~\ref{eq:twocc}:
\be
{1\over 4\pi}{\rm Im}\,\left[ \df\!\!\dtt \tr \pp (\Psi )
\right] \quad ,
\label{eq:ntwop}
\ee
where the holomorphic functional $\pp (\Psi )$ is
called {\bf the N=2 prepotential}.
Obviously
\be
\pp (\Psi ) = \half\tau\Psi ^2
\ee
gives back the classical N=2 Yang-Mills action of
Eq.~\ref{eq:ntwoact}.

Let us define
\bea
\pp _a(\Phi ) &=& {\partial\pp (\Phi )\over\partial\Phi _a}
\quad ,\nonumber\\
\pp _{ab}(\Phi ) &=& {\partial ^2\pp (\Phi )\over
\partial\Phi _a\partial\Phi _b}
\quad .
\eea
Then the general Lagrangian can be written in terms of
N=1 superfields as follows:
\be
{1\over 4\pi}{\rm Im}\,\left[ \dt \half\pp _{ab}(\Phi )W ^{\alpha a}
W^b_{\alpha} + \dtf (\Phi ^{\dagger}\e{2V} )^a\pp _a (\Phi )
\right]
\quad .
\ee
Thus from the N=1 point of view we have a special case of
Eq.~\ref{eq:mostgenact}: the superpotential vanishes,
the \Kahler\ potential is
\be
K = {\rm Im}\left[ (\Phi ^{\dagger}\e{2V} )^a\pp _a (\Phi )
\right] \quad ,
\ee
and the gauge kinetic function is
\be
f(\Phi ) = \pp _{ab}T^aT^b
\quad .
\ee

Notice that in this more general N=2 action the scalar
fields $A^a$ describe a nonlinear sigma model whose
target space \Kahler\ potential has the special form
above, i.e. it can be written in terms of a derivative of
a holomorphic function. The target space is a special
\Kahler\ manifold known as
the ``special \Kahler '' manifold. \cite{andy}

\subsection{N=2 hypermultiplets}

While $\Psi $ was assumed to be a singlet under the global $SU(2)$
symmetry, we can also consider a general N=2 scalar superfield
which is an $SU(2)$ doublet:
\[
\Phi ^a\sxt
\]
An N=2 hypermultiplet superfield is then defined by
the covariant constraints
\bea
D^a_{\alpha}\Phi ^b &=& \half\epsilon ^{ab}
D^c_{\alpha}\Phi ^c
\quad ,\nonumber\\
\barD ^a_{\alphadot}\Phi ^b &=& \half\epsilon ^{ab}
\barD ^c_{\alphadot}\Phi ^c
\quad .
\eea
These constraints simply remove the isotriplet parts of
$D^a_{\alpha}\Phi ^b$ and $\barD ^a_{\alphadot}\Phi ^b$:
\be
[\half ] + [\half ] = [0]_{\rm antisymm.} + [1]_{\rm symm.}
\quad .
\ee
The independent component fields are:
\begin{eqnarray*}
&A^a(x)\quad ,\quad &{\rm complex\ scalar\ isodoublet} \\
&\psi _{\alpha}(x),\;\chi _{\alpha}(x)\quad ,\quad &
{\rm two\ isosinglet\ spinors} \\
&F^a(x)\quad ,\quad &{\rm complex\ auxiliary\ scalar\ isodoublet}
\end{eqnarray*}
On-shell this implies $4$$+$$4=8$ real components, which as we
have already noted is twice the number in the massless N=2
hypermultiplet of single particle states.

A free superspace action for an N=2 hypermultiplet superfield
$\Phi ^a$ is
\be
\df D^{\alpha a}D^b_{\alpha}
\left[ \hbox{$\Phi ^a$}^{\dagger}
D^{\beta c}D_{\beta}^c\Phi ^b \right]
\quad .
\ee
With more difficulty, we can couple N=2 hypermultiplets
to N=2 Yang-Mills; the details are not particularly illuminating.

Note that there can be no renormalizable self-interaction
for $\Phi ^a$ since there is no cubic $SU(2)$ invariant.

We can construct N=2 generalizations of nonlinear sigma models
out of the hypermultiplets. It is easiest to start with the
N=1 case in components
\be
g_{ij*}\, \partial _m A^i(x) \partial ^mA^{*j} + \ldots \quad,
\ee
then impose the extra constraints of N=2 supersymmetry.
The end result \cite{luis} is that the target spaces of
N=2 hypermultiplet nonlinear sigma models are
{\bf hyper\Kahler} manifolds.

\section{Supergravity}

So far we have only considered global supersymmetry, generated
by
\[
\xi Q + \barxi\barQ
\]
with $\xi ^{\alpha}$, $\barxi _{\alphadot}$ {\bf constant}
Grassmann parameters. If we want {\bf local supersymmetry},
we should promote these parameters to functions of spacetime:
\be
\xi ^{\alpha},\;\barxi _{\alphadot}\rightarrow
\xi ^{\alpha}(x),\;\barxi _{\alphadot}(x) \quad .
\ee
Rigid superspace then becomes {\bf curved superspace}. From
the superspace vielbein $E_M^A$ and spin connection $W_A^{mn}$
we can construct the superspace curvature and torsion:
\[
\hbox{$R_{MNA}$}^B,\quad \hbox{$T_{MN}$}^A \quad .
\]

Recall that N=1 rigid superspace has already nonzero torsion,
so we {\bf cannot} constrain
all components of the curved superspace torsion to vanish
as we do in general relativity. On the other hand the superspace
vielbein and connection have too many independent components to
define a sensible theory. Thus the main difficulty in constructing
supergravity theories is finding and solving an appropriate set
of covariant constraints. This gets very complicated, \cite{sugra}
and is beyond the scope of these lectures.

Let us instead quote results. One can construct an off-shell
supergravity multiplet with the following field content:
\begin{eqnarray*}
&e_m^a \;,\quad &{\rm vierbein\rightarrow spin\ 2\ graviton}\\
&\psi _m^{\alpha} \;,\quad &
{\rm vector\hbox{-}spinor\rightarrow spin\ 3/2\ gravitino}\\
&b_a \;,\quad &{\rm auxiliary\ real\ vector\ field}\\
&M \;,\quad &{\rm auxiliary\ complex\ scalar\ field}
\end{eqnarray*}

{}From these fields we can construct a supergravity Lagrangian:
\bea
\lefteqn{8\pi G\lag = -\half eR -\third e|M|^2 +\third eb^ab_a }
\nonumber\\
& & +\half e \epsilon ^{klmn}\left( \barpsi _k\barsigma _l
{\cal D}_m\psi _n - \psi _k\sigma _l{\cal D}_m\barpsi _n \right)
\eea
where:
\begin{eqnarray*}
G &=& {\rm Newtons\ constant,}
\\
e &=& {\rm det}\, e^a_m,
\\
R &=& {\rm Ricci\ scalar\ curvature,}
\\
{\cal D}_m &=& {\rm covariant\ derivative\ for\ spin\ 3/2\ fields.}
\end{eqnarray*}
The action is invariant under
\begin{itemize}
\item general coordinate transformations,
\item local Lorentz transformations,
\item local N=1 supersymmetry.
\end{itemize}

Let's count the off-shell degrees of freedom of N=1 supergravity.
Because the action is invariant under three types of local symmetries
we should only count gauge invariant degrees of freedom:
\begin{eqnarray*}
&e^a_m \;:\; &4\times 4 = 16 \\
& & - 4 {\rm\ general\ coordinate\ }\zeta ^m \\
& & - 6 {\rm\ local\ Lorentz\ }\lambda ^{ab} \\
& & = 6 {\rm\ bosonic\ real\ components} \\
& & \\
&\psi _m^{\alpha} \;:\; &4\times 4 = 16 \\
& & - 4 {\rm\ local\ N=1\ SUSY\ }\xi ^{\alpha} \\
& & = 12 {\rm\ fermionic\ real\ components} \\
& & \\
&b_a \;:\; &{\rm\ real\ vector} \\
& & = 4 {\rm\ bosonic\ real\ components} \\
& & \\
&M \;:\; &{\rm\ complex\ scalar} \\
& & = 2 {\rm\ bosonic\ real\ components}
\end{eqnarray*}
Thus we have a total of 12 bosonic and 12 fermionic real
components in the off-shell N=1 supergravity multiplet.
On-shell we have only $2 + 2$ components, corresponding to
a massless spin 2 graviton and a massless spin 3/2 gravitino.

Of course, we really want to be able to couple N=1 supergravity
to N=1 supersymmetric Yang-Mills and N=1 chiral superfield matter,
all in a way which is consistent with local supersymmetry.
This is again a complicated problem and the final result is
not particularly intuitive. \cite{wb}

We can also extend 4-dimensional N=1 supergravity
to N=2, 3, 4, or 8 supergravity. These extended supergravities
automatically couple gravity to gauge fields and matter fields
in a way consistent with local supersymmetry, just as N=2
Yang-Mills couples gauge fields to matter in a way consistent
with global supersymmetry. Extended supergravities are easier
to construct and understand if we use dimensional reduction.
For example, 4-dimensional N=8 supergravity can be obtained by
dimensionally reducing 11-dimensional N=1 supergravity, which
is a rather simple theory to describe. We will return to this
fact when we discuss supersymmetry in higher dimensions.

\section{Renormalization of N=1 SUSY Theories}

Consider again the Wess-Zumino model:
\be
\dff \Phi ^{\dagger}\Phi -
\df \left[ \dt (\half m\Phi ^2 + \third g\Phi ^3 ) +{\rm h.c.}
\right] \quad ,
\ee
We would like to work out the superfield Feynman rules of this theory.
However we encounter an
immediate difficulty which is that $\Phi$ is not a
general scalar superfield, but rather a constrained superfield.
Thus in computing perturbative diagrams with chiral superfields
we must deal with the occurence of integrals $\dft$ over only
part of the full N=1 rigid superspace.

This difficulty is overcome by
introducing a {\bf projection operator} for chiral superfields.
The projection operator we need is
\be
P_+ \equiv -{1\over 16\BBox }\barD ^2D^2\quad .
\ee
This operator clearly has the property that it
takes a general scalar superfield to a chiral superfield, i.e.
\be
\barD _{\alphadot}P_+\Phi\sx = 0
\ee
follows trivially from the fact that $(\barD )^3 = 0$.
To prove that $P_+$ is in fact a projection operator,
we must also show that
\[
(P_+)^2 = P_+ \quad .
\]
We use the following identity (which can be verified by
brute force):
\be
{[}\,\barD^2,D ^2\, {]} =
8i(D\sigma ^m\barD )\partial _m - 16\BBox \quad .
\ee
Thus:
\bea
(P_+)^2 &=& {1\over 16\BBox }\barD ^2D^2{1\over 16\BBox }\barD ^2D^2
\nonumber\\
&=& \left( {1\over 16\BBox } \right)^2 \barD ^2D^2\barD ^2D^2
\nonumber\\
&=& \left( {1\over 16\BBox } \right)^2 \barD ^2D^2
(D^2\barD ^2 + 8i(D\sigma ^m\barD )\partial _m - 16\BBox )
\\
&=& \left( {1\over 16\BBox } \right)^2 \barD ^2D^2
(-16\BBox )
\nonumber\\
&=& P_+ \quad ,
\eea
where in the fourth line we used $(D)^4$$=$$(D)^3$$=$$0$.

Similarly the projection operator for antichiral superfields is
\be
P_- \equiv -{1\over 16\BBox }D^2\barD ^2\quad .
\ee

We can now deal with
the occurence of integrals $\dft$ over only
part of the full N=1 rigid superspace. The judicious use of
projection operator insertions allow us to convert
these into integrals over the entire superspace.
For example:
\bea
\dt m\, \Phi\Phi &=& \dt m\, \Phi P_+ \Phi \nonumber\\
&=& -4\dtf m\, \Phi {1\over 16\BBox}D^2 \Phi
\quad ,
\eea
where in the last line we used the fact that
$\barD ^2\Phi = 0$ and that, modulo surface terms,
\be
\df\barD ^2 \equiv -4\df\!\!\int d^2\barth \quad .
\ee

A similar difficulty occurs for the cubic interaction
term $g\Phi ^3$. These vertices correspond to
functional derivatives with respect to chiral
sources $J\sy$. These functional derivatives produce
superspace delta functions
\[
\delta ^4(x_1 - x_2)\delta ^2(\th _1 - \th _2)
\]
whereas what we want (for internal lines anyway)
are delta functions for the full superspace:
\bea
\delta ^4(x_1 - x_2)\delta ^4(\th _1 - \th _2)
&=&  \delta ^4(x_1 - x_2)\delta ^2(\th _1 - \th _2)
\delta ^2(\barth _1 - \barth _2)
\nonumber\\
&=& \delta ^4(x_1 - x_2)(\th _1 - \th _2)^2(\barth _1 - \barth _2)^2
\eea
This is easily remedied by using the identity
\be
\delta ^4(x_1 - x_2)\delta ^2(\th _1 - \th _2)
= -\quarter \barD ^2
\delta ^4(x_1 - x_2)\delta ^4(\th _1 - \th _2)
\quad .
\ee
In loop graphs, one factor of $\barD ^2$
from each vertex will get used
up converting an $\dt$ to an $\dtf$.
Of course we also have a similar trick for the
$g\hbox{$\Phi ^{\dagger}$}^3$ vertices.

We can now compute the superspace propagator of
the Wess-Zumino model by performing the functional integral
of the quadratic part of the action, written in the form:
\bea
&&\dff \Biggl[ \half (\Phi\;\Phi ^{\dagger})
\left( \matrix{\quarter{m\over\littlebox}D^2&1\cr
                       1&\quarter{m\over\littlebox}\barD ^2\cr }
\right)\left( {\Phi \atop \Phi ^{\dagger}} \right)
\nonumber\\
&&\qquad\qquad + (\Phi\;\Phi ^{\dagger})\left(
{-\quarter{D^2\over\littlebox}J\atop -\quarter{\barD ^2\over\littlebox}\barJ}
\right) \Biggr]
\quad .
\eea

\subsection{Nonrenormalization}

Without further ado we can now summarize {\bf the superspace
dependence} of the resulting Feynman rules for 1PI diagrams:
\begin{itemize}
\item There is an $\dtf$ for each vertex.
\item For a $\Phi ^3$ vertex $n$ of whose lines are external,
there are $2$$-$$n$ factors of $\barD ^2$. For a
$\hbox{$\Phi ^{\dagger}$}^3$ vertex, there are
$2$$-$$n$ factors of $D ^2$.
\item There is a Grassmann delta function
\be
\delta ^4(\th _1 -\th _2) = (\th _1 -\th _2)^2(\barth _1 - \barth _2)^2
\ee
for each propagator.
\item There is a factor of $D^2$ for each $\Phi$-$\Phi$
propagator, and a factor of $\barD ^2$ for each
$\Phi ^{\dagger}$-$\Phi ^{\dagger}$ propagator.
\end{itemize}

Consider now an arbitrary loop graph. Integrating
the various $D^2$ and $\barD  ^2$ factors by parts, we
can perform all but one of the $\dtf$ integrations using
the delta functions. Let $\int d^4\th$ denote the
final integration, and $\delta ^4(\th -\th _2)$ be
the one remaining Grassmann delta function. This delta
function is already supposed to be evaluated at $\th = \th _2$,
due to the $\th _2$ integral already performed. So the graph
vanishes unless there is precisely one factor of $D^2$ and
one factor of $\barD ^2$ acting on the final delta function:
\[
D^2\barD ^2\delta ^4(\th - \th_2) = 16 \quad .
\]

The only remaining $\th$ dependence comes from the external lines.
Thus an arbitrary term in the effective action can be reduced
to the form:
\be
\int d^4\th\!\!\int d^4x_1\cdots d^4x_n\; F_1(x_1,\th ,\barth )
\cdots F_n(x_n,\th ,\barth )G(x_1,\ldots x_n)
\quad ,
\ee
where the $F$'s are superfields and covariant derivatives of
superfields, and all the spacetime structure is swept into
the translationally invariant function $G(x_1,\ldots x_n)$.

This result is called the {\bf N=1 Nonrenormalization Theorem}.
It generalizes to N=1 actions containing
arbitrary numbers of chiral and vector superfields.
An important consequence is that if all of the external lines are chiral,
or if all of the external lines are antichiral, the expression
above vanishes. Thus:
\vskip .2in{\bf \begin{center}
The superpotential is not renormalized at any order in
perturbation theory.
\end{center}}\vskip .2in

Another important result of the above analysis is that
all vacuum diagrams and tadpoles vanish. This is consistent with the
fact that the vacuum energy is precisely zero in any
globally supersymmetric theory.

Note our derivation implicitly assumed that the spacetime
loop integrals are regulated in a way which is consistent
with supersymmetry. This is {\bf not} the case if we employ
dimensional regularization, since the numbers of fermions
and bosons vary differently as you vary the dimension.
Supersymmetric loop diagrams are usually evaluated using
a regularization called {\bf dimensional reduction},
where the spinor algebra is fixed at four dimensions while
momentum integrals are performed in $4-2\epsilon$ dimensions.
This is not a completely satisfactory procedure either. \cite{west}

\subsection{Renormalization}

What renormalization do we have to perform in
an N=1 SUSY model with chiral superfields $\Phi ^i$
and a vector superfield $V$? We have wave function
renormalization:
\bea
\Phi ^i_0 &=& \hbox{$Z^{1/2}$}^{ij}\,\Phi ^j \quad ,
\nonumber\\
V_0 &=& Z_V^{1/2}\, V \quad ,
\eea
and we also have gauge coupling renormalization:
\be
g_0 = Z_g\, g \quad .
\ee
Even better, if we compute using the background field
method, the background field gauge invariance implies
the relation: \cite{west}
\be
Z_g\, Z_V^{1/2} = 1 \quad .
\ee

The end result is that we can characterize the renormalized
theory in terms of two objects:
\begin{itemize}
\item The beta function:
\[
\beta (g) = \mu {\partial\over\partial\mu}g \quad ;
\]
\item The anomalous dimensions matrix of the $\Phi ^i$:
\[
\gamma ^{ij} = \hbox{$Z^{-1/2}$}^{ik}\mu
{\partial\over\partial\mu}\hbox{$Z^{1/2}$}^{kj} \quad .
\]
\end{itemize}

\section{Holomorphy and the N=2 Yang-Mills Beta Function}

In this section we will review some of Seiberg's original
arguments about the N=2 supersymmetric $SU(2)$ Yang-Mills
beta function. \cite{oldsei} This type of argumentation
deals with the effective infrared (i.e. low energy) limit
of the theory, described by the {\bf Wilsonian action}. \cite{shif}
The form of this effective action will be constrained by
three kinds of considerations:
\begin{itemize}
\item holomorphy,
\item global symmetries,
\item the existence of a nonsingular weak coupling limit.
\end{itemize}

Let us begin by listing the global symmetries of
the classical N=2 supersymmetric $SU(2)$ Yang-Mills Lagrangian.
These are: the global $SU(2)$ R symmetry arising from the
automorphism of the N=2 algebra, and an additional $U(1)$ R
symmetry defined by:
\bea
\th &\rightarrow & \e{i\omega}\th \quad ,\nonumber\\
\tht &\rightarrow & \e{i\omega}\tht \quad ,\\
\Psi &\rightarrow & \e{2i\omega}\Psi \quad .
\eea
There is an axial current $j^R_m$ corresponding to
the $U(1)$ R symmetry. Since both fermions $\psi$ and
$\lambda$ have R-charge +1, there is a nonvanishing
ABJ triangle anomaly. We write the anomalous divergence of the
R current, remembering that the fermions are
in the adjoint representation:
\be
\partial ^mj^R_m = 2C_2(G){g^2\over 16\pi ^2} F_{mn}
{\tilde{F}}^{mn} = {g^2\over 4\pi ^2} F_{mn}{\tilde{F}}^{mn}
\quad .
\label{eq:anom}
\ee

Next we deduce the {\bf moduli space} of gauge inequivalent
classical vacua for N=2 supersymmetric $SU(2)$ Yang-Mills.
The theory contains an $SU(2)$ triplet complex scalar field
$A(x)$ whose scalar potential is (see Eq.~\ref{eq:vd}):
\be
V(A) = {1\over 2g^2}D^2 =
{g^2\over 8}\left( [A,A^*] \right)^2\quad .
\ee
Unbroken supersymmetry requires that $V(A) = 0$ in the vacuum.
Up to a gauge transformation, the most general solution to
this requirement is:
\be
A(x) = \half a \sigma ^3 \quad ,
\ee
where $\sigma ^3$ is the Pauli matrix and $a$ is a complex constant.
The parameter $a$ does not quite give only gauge inequivalent vacua,
since by Weyl symmetry vacua labeled $a$ and $-a$ are gauge
equivalent. So the classical moduli space is described by a complex
parameter $u$, with
\be
u = \half a^2 = \langle \tr A^2 \rangle \quad .
\ee

For a generic nonvanishing value of $u$, the $SU(2)$ gauge
symmetry is broken down to a $U(1)$. Since N=2 SUSY is still
in force, masses are generated not just for two components of
the $SU(2)$ gauge field, but also for their N=2 superpartners.
Thus the remaining light fields consist of a $U(1)$ gauge boson,
a massless uncharged complex scalar, and two massless uncharged
fermions. The infrared effective action clearly exists in this
case. The point $u=0$, on the other hand, appears singular.

The form of the infrared effective action is severely constrained
by N=2 supersymmetry. The part of this action which contains no
more than two spacetime derivatives and interactions of no more
than four fermions must have the same form as the classical action
of the ultraviolet theory:
\be
{1\over 4\pi}{\rm Im}\,\left[ \df\!\!\dtt \tr \ppf (\Psi )
\right] \quad ,
\label{eq:ntwopf}
\ee

Because of the anomaly, the effective action is not
invariant under a $U(1)_R$ transformation. Instead, the
Adler-Bardeen theorem tells us that the effective action is shifted by
\bea
&&\omega \df {1\over 4\pi ^2} F_{mn}{\tilde{F}}^{mn} \nonumber\\
=&& -\omega\; {\rm Im}\;\left[ \int d^4x\!\!\dtt {1\over 2\pi ^2}
\,\Psi ^2 \right]
\quad .
\label{eq:shift}
\eea
Neglecting (for the moment) instanton effects, the effective
action is still constrained by $U(1)_R$ invariance modulo
this shift, which is manifestly a one-loop effect.
The only holomorphic functional of $\Psi$ with this property
is
\be
\pp _1 =
i{1\over 2\pi}\Psi ^2\,{\rm ln}{\Psi ^2\over\Lambda ^2} \quad ,
\ee
where $\Lambda$ is a dynamically generated scale. Actually,
since we can absorb the tree-level $\pp$ into the definition
of $\Lambda$, $\pp _1$ is the full effective prepotential
to all orders in perturbation theory.

{}From the shift Eq.~\ref{eq:shift} we see that a single instanton
violates R charge conservation by $8$ units, breaking the global
$U(1)_R$ symmetry down to $Z_8$ (in fact, since $u$ carries
R charge 4, there is a further breaking down to $Z_4$).
This suggests that the
complete nonperturbative effective prepotential has the form:
\be
\pp = i{1\over 2\pi}\Psi ^2\,{\rm ln}{\Psi ^2\over\Lambda ^2}
+ \sum _{k=1}^{\infty} \pp _k \,\left( {\Lambda\over\Psi} \right)^{4k}
\Psi ^2
\quad ,
\label{eq:fullef}
\ee
where the $\pp _k$ are numerical coefficients, and the
$k$th term arises as a contribution of $k$ instantons.

Returning to the all orders perturbative result, we can
deduce the effective Wilsonian gauge coupling $g_{\rm W}(u)$ from the gauge
kinetic function:
\be
f(\Phi ) = {\delta ^2\pp (\Phi )\over\delta\Phi\delta\Phi}
\quad ;
\ee
thus:
\be
{1\over g_{\rm W}(u)} =
{1\over g^2_0}\left[ 1 + {3g^2_0\over 4\pi ^2}
+ {g^2_0\over 4\pi}{\rm ln}\left({u\over\Lambda ^2}\right)
\right] \quad .
\ee
This gives us the all-orders perturbative beta function:
\be
\beta (g) = -{g^3\over 4\pi ^2}
\quad .
\label{eq:allord}
\ee

We could try to make an analogous derivation in the case
of N=1 SUSY Yang-Mills. However in the N=1 case there is a
separate wave function renormalization of the D-term of the
chiral superfield action. Because of this $g_{\rm W}$$\ne$$g_{\rm eff}$
and Eq.~\ref{eq:allord} is not valid.

{}From the expression for the effective gauge coupling
we see that the $u\rightarrow\infty$ limit is the
weak coupling limit. This explains why we did not
include negative $k$ contributions to
Eq.~\ref{eq:fullef}, i.e. why the sum extends only
over positive $k$. Terms with negative $k$ blow up like a power
as $u\rightarrow\infty$, behavior which is inconsistent
with the existence of a nonsingular weak coupling limit.

\section{Supersymmetry in spacetime dimensions 2, 6, 10, and 11}

\subsection{Spinors in arbitrary spacetime dimensions}

The dimension of Dirac spinors in $d$ spacetime dimensions
can be deduced by constructing the Dirac gamma matrices
obeying the Clifford algebra
\be
\{ \gamma ^m,\gamma ^n \} = 2\eta ^{mn}
\quad .
\ee
The result is
\be
d_{\gamma} = \cases{ 2^{d/2}&d{\rm\ even,}\cr
                     2^{(d-1)/2}&d{\rm\ odd.}\cr}
\ee

Starting with Dirac spinors, we can investigate whether
it is possible to impose Weyl, Majorana, or simultaneously
Weyl and Majorana conditions on these spinors.

For Weyl spinors, we need to generalize the notion of
the chirality operator $\gamma ^5$. Recall that in four
dimensions, CPT conjugate spinors have opposite chirality,
implying that there are no gravitational anomalies. \cite{alv}
This is related to the fact that $\gamma ^5$, defined as
\be
\gamma ^5 = \gamma ^0\gamma ^1\gamma ^2\gamma ^3
\ee
has eigenvalues $\pm i$, since
\be
(\gamma ^5)^2 = -I
\quad .
\label{eq:gasq}
\ee
For any spacetime dimension $d = 4k$, $k=1,2,\ldots$,
we can define ``$\gamma ^5$'' in an exactly analogous way:
\be
\gamma ^5 = \gamma ^0\gamma ^1\cdots\gamma ^{4k-1}
\quad ,
\ee
and Eq.~\ref{eq:gasq} still holds. Thus in $d=4k$ dimensions
Weyl spinors exist and gravitational anomalies are absent.
In $d=4k$$+$$2$ dimensions
\be
(\gamma ^5)^2 = I
\quad ,
\ee
implying that CPT conjugate spinors have the same chirality.
Thus Weyl spinors exist and gravitational anomalies are
possible. In odd dimensions $\gamma ^5$ defined as above
is the identity; there is no chirality operator and thus
no Weyl spinors.

The analysis of Majorana and Majorana-Weyl conditions in
arbitrary dimensions is more involved; a good reference
is Sohnius. \cite{soh} The results are summarized
in Table~\ref{tab:spinors}.

\begin{table}[t]
\caption{Properties of spinors in spacetime dimensions 2 to 12.
$d_{\gamma}$ is the dimension of Dirac gamma matrices.
\label{tab:spinors}}

\vspace{0.4cm}
\begin{center}
\begin{tabular}{|c|c|c|c|c|c|c|c|c|c|c|c|}
\hline
$d$&2&3&4&5&6&7&8&9&10&11&12\\
\hline
&&&&&&&&&&&\\
$d_{\gamma}$&2&2&4&4&8&8&16&16&32&32&64\\
\hline
minimum spinor dim.&1&2&4&8&8&16&16&16&16&32&64\\
\hline
&&&&&&&&&&&\\
Weyl?&X&&X&&X&&X&&X&&X\\
\hline
Majorana?&X&X&X&&&&X&X&X&X&X\\
\hline
Majorana-Weyl?&X&&&&&&&&X&&\\
\hline
&&&&&&&&&&&\\
gravitational anomalies?&X&&&&X&&&&X&&\\
\hline
\end{tabular}
\end{center}
\end{table}

\subsection{Supersymmetry in arbitrary spacetime dimensions}

To discuss supersymmetry in spacetime dimensions other than
four, we need an improved notation for keeping track of
the number of independent supersymmetry generators.
In four dimensions N=1 SUSY means that there are four
independent SUSY generators:
\[
Q_1,\; Q_2,\; \barQ _{\onedot},\; \barQ _{\twodot} \quad .
\]
This is of course the minimum number of supersymmetries in
four dimensions since the minimum spinor dimension is four.
Let us refer to this as N=(1)$_4$ supersymmetry. More generally,
N=(p)$_{p*n_{\rm min}}$ denotes $p*n_{\rm min}$ supersymmetries,
where $n_{\rm min}$ is the minimum spinor dimension.

In this new notation the possible global supersymmetries in
four dimensions are:
\begin{eqnarray*}
N &=& (1)_4 \\
N &=& (2)_8 \\
N &=& (4)_{16} \\
N &=& (8)_{32} \\
&\ldots &
\end{eqnarray*}

In $4k$$+$$2$ dimensions we can have {\bf independent}
chiral and antichiral SUSY generators, since CPT conjugates
have the same chirality. Thus we need a notation which
distinguishes chiral from antichiral SUSY generators:
\[
N = (p,q)_{(p+q)*n_{\rm min}}
\]
where $p$, $q$ are the number of chiral/antichiral SUSY
generators, respectively.

{}From Table~\ref{tab:spinors} it is clear that in any
spacetime dimension there is a {\bf minimum} number
of supersymmetries (other than zero). Thus:

\vskip .2in
\begin{flushleft}
As few as \ 4 supersymmetries can only occur for:\leaderfill $d\le 4$\\
\ \\
As few as \ 8 supersymmetries can only occur for:\leaderfill $d\le 6$\\
\ \\
As few as 16 supersymmetries can only occur for:\leaderfill $d\le 10$\\
\ \\
As few as 32 supersymmetries can only occur for:\leaderfill $d\le 11$\\
\end{flushleft}
\vskip .2in

Furthermore, in any spacetime dimension, the {\bf maximum}
number of supersymmetries of physical interest is always 32 or less.
This is because for more than 32 supersymmetries all
massless multiplets contain unphysical higher spin particles,
i.e. particles with spin greater than that of the $d$-dimensional
graviton.

Clearly we expect that many SUSY theories in
different spacetime dimensions but with the same number
of supersymmetries can be related, presumably through some
form of dimensional reduction or truncation. Indeed this
is true as we will see in several examples. Of particular
interest is the possibility of relating models with extended
supersymmetry in four dimensions to simpler models in the
``mother'' dimensions 6, 10, and 11.

\subsection{Supersymmetry in 2 dimensions}

In two dimensions we have $(p,q)$ type superalgebras.
I will briefly describe some examples.
\begin{itemize}
\item $(1,0)_1$ supersymmetry: here we have a single
left-handed Majorana-Weyl spinor:
\bea
Q_+ &=& Q_+^{\dagger} \quad ,\nonumber\\
\{ Q_+,Q_+ \} &=& 2iP_z \quad ,\\
{[}\,Q_+,P_z\,{]} &=& {[}\, Q_+,P_{\bar{z}}\,{]} = 0 \quad ,
\eea
where the antihermitian generators
$P_z$, $P_{\bar{z}}$ generate left and right-moving
translations in the two-dimensional spacetime parameterized
by coordinates $z$, $\bar{z} = x^0 \pm x^1$.
The minimal SUSY multiplet has just two states: one left-moving
real scalar (a ``chiral boson''), and one left-moving
Majorana-Weyl fermion.
\item $(1,1)_2$ supersymmetry: here we can construct a
$(1,1)$ supersymmetric nonlinear sigma model by supersymmetrizing
\be
\int dzd\bar{z}\, g_{ij}(A)\partial _zA^i\partial _{\bar{z}}A^j
\quad .
\ee
The target spaces of such models are general Riemannian manifolds.
\item $(2,2)_4$ supersymmetry: here again we can construct
a supersymmetric nonlinear sigma model. The target spaces are
\Kahler\ manifolds. Note that this agrees with the four-dimensional
N=(1)$_4$ case, which has the same number of supersymmetries.
\item $(4,4)_8$ supersymmetry: here we have supersymmetric
nonlinear sigma models whose target spaces are hyper\Kahler\ manifolds.
Again this agrees with the four-dimensional N=(2)$_8$ case,
which has the same number of supersymmetries.
\end{itemize}

Two-dimensional supersymmetry has important applications to
superstring theory, where it is interpreted as {\bf worldsheet
supersymmetry} rather than spacetime SUSY. For more details
see Hirosi Ooguri's lectures in this volume.

\subsection{Supersymmetry in 6 dimensions}

In six dimensions the minimal case is N=(1,0)$_8$ or
(0,1)$_8$. The SUSY generators can be expressed as a
single complex Weyl spinor:
\bea
Q_a\;&,&\quad a=1,\ldots 8; \nonumber\\
\{ Q_a,\barQ ^b \} &=& \half (1+\gamma ^7)_a^c(\gamma ^M)_c^b
P_M
\quad ,
\label{eq:sixsusy}
\eea
where we use capital Roman letters to denote 6-dimensional
spacetime indices, and $\gamma ^7$ is the chirality
operator, i.e. the 6-dimensional version of ``$\gamma ^5$''.

In six dimensions massless particles are labelled by
irreps of the little group $Spin(4) \sim SU(2)$$\times$$SU(2)$.
Let us describe the possible massless irreps of (0,1)$_8$
supersymmetry in terms of their $SU(2)$$\times$$SU(2)$ ``helicity''
states:
\begin{itemize}
\item {\bf hypermultiplet:} $2(\half ,0) + 4(0,0)$, i.e.
one complex Weyl fermion and two complex scalars, for a total
of $4$$+$$4 = 8$ real components.
\item {\bf vector multiplet:} $(\half ,\half ) + 2(0,\half )$,
i.e. a massless vector and one complex anti-Weyl fermion,
for a total of $4$$+$$4 = 8$ real components.
\item {\bf gravity multiplet:} $(1,1) + 2(\half ,1) + (0,1)$,
i.e. a graviton, two gravitini, and one self-dual 2nd rank
antisymmetric tensor, for a total of $9$$+$$12$$+3 = 24$
real components.
\item {\bf tensor multiplet:} $(1,0) + 2(\half ,0) + (0,0)$,
i.e. one anti-self-dual 2nd rank antisymmetric tensor, one
complex Weyl fermion, and one real scalar ``dilaton'', for a
total of $3$$+$$4$$+$$1 = 8$ real components.
\end{itemize}

\subsubsection{Dimensional reduction 6 $\rightarrow$ 4}

Consider six-dimensional N=(0,1)$_8$ supersymmetric
Yang-Mills theory. We write the action in terms of
fields describing the on-shell massless vector multiplet
described above. The action is:
\be
\int d^6x\, \left[ -\quarter F_{MN}F^{MN} +
\half i \barlam\gamma ^M\grad{M}\lambda  \right]
\quad,
\ee
where $\grad{M}$ is a gauge covariant derivative.

Now we imagine compactifying this theory on a torus.
Let $x^4$, $x^5$ be the compactified coordinates.
Then the 6-dimensional gauge field $A_M$ breaks up into
a 4-dimensional gauge field $A_m$ and two real scalars
$A_4$, $A_5$. The complex anti-Weyl fermion $\lambda$
breaks up into two 4-dimensional complex Weyl fermions.
In addition we will have an infinite tower of massive
Kaluza-Klein states, corresponding to the Fourier
decomposition
\be
\phi (x^m,x^4,x^5) = \sum _{n_4,n_5}\,
\e{[-in_4m_4x^4-in_5m_5x^5]}\phi _{n_4n_5}(x^m)
\quad,
\ee
where $\phi (x^m,x^4,x^5)$ denotes any 6-dimensional field
component, and $m_4$, $m_5$ are inversely related to
the compactification radii.

The SUSY generator $Q_a$ splits up into two
$Q_{\alpha}$'s, implying that the 4-dimensional theory
has N=2 supersymmetry. The 6-dimensional translation
generator $P_M$ splits into $P_m$, $P_4$, $P_5$.
{}From the 6-dimensional SUSY algebra we see that $P_4$ and $P_5$
commute with the $Q_{\alpha}$'s and with $P_m$. They also
appear on the right-hand side of $\{Q,Q\}$. Clearly what
we have here are two real $=$ one complex central charge:
\[
X^{ab} \sim (P_4+iP_5)\epsilon ^{ab} \quad .
\]
Multiplying the two $Q_{\alpha}$s by a phase to convert
to Zumino's basis (see Eq.~\ref{eq:zurel}), this corresponds
to a single real central charge
\be
Z = \sqrt{P_4^2 + P_5^2} \quad .
\ee
Thus the dimensionally reduced theory consists of 4-dimensional
N=2 supersymmetric Yang-Mills with central charges and an
infinite number of massive multiplets.

Now suppose we repeat the above excercise, but first adding
some massless $d=6$ hypermultiplets $(A,B,\psi )$
to our N=(0,1)$_8$ supersymmetric
Yang-Mills theory. Each $d=6$ hypermultiplet will give one
$d=4$ multiplet of fields with the counting of the
$d=4$ N=2 massless hypermultiplet, plus a tower of additional
massive multiplets. The 6-dimensional on-shell condition for the
complex scalars $A$ and $B$
\be
\partial ^M\partial _M A = \partial ^M\partial _MB = 0
\ee
fixes the masses of the 4-dimensional hypermultiplet scalars:
\be
4m^2 = Z^2 \quad .
\ee
Thus the dimensionally reduced theory contains
massive BPS-saturated N=2 short multiplets, which as already
noted do indeed have the same counting as the
$d=4$ N=2 massless hypermultiplet.

\subsubsection{Anomaly cancellation}

A striking feature of the 6-dimensional gravity multiplet
is that it contains the self-dual part of a 2nd rank
antisymmetric tensor, without the anti-self-dual part.
As is also the case in four dimensions, it is impossible
to write a Lorentz covariant Lagrangian formulation of just
the self-dual antisymmetric tensor field. However one can
write Lorentz covariant equations of motion, and it appears
that the corresponding field theory exists and is Lorentz
invariant, despite the lack of a {\bf manifestly} Lorentz
invariant action principle. If
\be
n_g = n_t = 1\quad ,
\ee
where $n_g$, $n_t$ are the number of gravity and tensor
multiplets, then of course we can write a Lorentz covariant
Lagrangian. Thus we have the interesting result that
every 6-dimensional supergravity theory {\bf either} contains
a dilaton field {\bf or} has no manifestly Lorentz invariant
action principle.

In six dimensions we have both gravitational anomalies
and mixed gauge-gravitational anomalies. Anomaly cancellation
is a severe constraint on the particle content, and in
particular on which combinations of SUSY multiplets yield
anomaly-free theories.

For example, when $n_g$$=$$n_t$$=$$1$, the necessary and
sufficient condition for anomaly cancellation is:
\be
n_h = n_v + 244 \quad ,
\label{eq:cancon}
\ee
where $n_h$, $n_v$ are the number of hypermultiplets and
vector multiplets. Thus we always need a remarkably large
number of hypermultiplets to cancel anomalies.

Let's look at two examples of anomaly-free
6-dimensional N=(0,1)$_8$ supersymmetric
supergravity-Yang-Mills-matter theories.
\begin{itemize}
\item Gauge group $E_8\times E_7$, with 10 massless hypermultiplets
in the $(1,56)$ of $E_8\times E_7$, and 65 singlet hypermultiplets.
Thus:
\bea
n_v &=& 248 + 133 = 381 \quad ;\nonumber\\
n_h &=& 560 + 65 = 625 \quad ;
\eea
which satisfies Eq.~\ref{eq:cancon}.
\item Gauge group $SO(28)\times SU(2)$, with 10 massless
hypermultiplets in the $(28,2)$ of $SO(28)\times SU(2)$,
and 65 singlet hypermultiplets. Thus:
\bea
n_v &=& 378 + 3 = 381 \quad ;\nonumber\\
n_h &=& 560 + 65 = 625 \quad ;
\eea
which satisfies Eq.~\ref{eq:cancon}.
\end{itemize}

These examples arise as compactifications of the ten-dimensional
$E_8\times E_8$ and $SO(32)$ heterotic strings, respectively,
onto the complex dimension 2 \Kahler\ manifold K3.

\subsection{Supersymmetry in 11 dimensions}

Eleven is the maximum dimension in which we can have
{\bf as few as} 32 supersymmetries. Thus $d=11$ is the
maximum dimension of interest to supersymmetry theorists,
unless one is willing to make some drastic assumptions
such as altering the Minkowski signature of spacetime.

Futhermore, there is only one sensible supersymmetric
theory in eleven dimensions: N=1 supergravity.
The N=(1)$_{32}$ SUSY algebra is generated by a single
Majorana spinor $Q_a$, $a=1,\ldots 32$.

Here is
the field content of the on-shell massless $d=11$ N=1 gravity multiplet,
characterized by irreps of the little group $SO(9)$:
\begin{itemize}
\item $e^A_M$, the 11-dimensional vielbein. On-shell this
constitutes a 44 of $SO(9)$.
\item $\psi ^a_M$, $a = 1,\ldots 32$, an 11-dimensional
massless vector-spinor, i.e. the gravitino field. On-shell this
is a 128 of $SO(9)$.
\item $A_{MNP}$, a 3rd rank antisymmetric tensor. On-shell
this is an 84 of $SO(9)$.
\end{itemize}

The Lagrangian of 11-dimensional supergravity, in terms of
these component fields of the on-shell multiplet, is rather
simple. \cite{julia}
The first three terms are:
\be
-{1\over 2\kappa ^2}eR -\half e\barpsi _M\Gamma ^{MNP}
D_N\psi _P - {1\over 48}eF_{MNPQ}F^{MNPQ}
\quad ,
\ee
where:
\begin{eqnarray*}
\kappa &=& {\rm 11\hbox{-}dimensional\ gravitational\ coupling}\quad ,\\
e &\equiv & {\rm det}\,e^A_M \quad ,\\
R &=& {\rm Ricci\ scalar}\quad ,\\
\Gamma ^{MNP} &=& e_A^Me_B^Ne_C^P\,
\gamma ^{[A}\gamma ^B\gamma ^{C]} \quad ,\\
D_N &=&
{\rm Lorentz\ covariant\ derivative\ for\ 11
\hbox{-}dim.\ vector\hbox{-}spinors}
\quad ,\\
F_{MNPQ} &=& \partial _{[M}A_{NPQ]},{\rm\ i.e.\ a\ field\ strength}
\quad ,
\end{eqnarray*}
and the square brackets denote antisymmetrization.

Eleven-dimensional supergravity is the field theory limit
of {\bf M-theory}, which is in turn a
strong coupling limit of superstring theory.

\subsubsection{Dimensional reductions of $d=11$ supergravity}

We can truncate $d=1$ supergravity down to a 4-dimensional
theory by simply suppressing the dependence on $x^4$-$x^{10}$.
Since the resulting 4-dimensional theory still has 32
supersymmetries, we obviously must get N=(8)$_{32}$ extended
supergravity. Provided one is satisfied with on-shell formulations,
this is a simpler way of deriving the
rather complicated 4-dimensional theory.

Another useful example is to truncate $d=11$ to $d=10$.
The Majorana spinor $Q_a$ (32 components) splits into two
10-dimensional Majorana-Weyl spinors $Q_{\alpha}$,
$\tilde{Q}_{\alphadot}$ (16 components each) with opposite
chirality. Thus the truncated theory is a nonchiral $d=10$
N=(1,1)$_{32}$ supergravity, commonly known as
{\bf Type IIA supergravity}. \cite{green}

The components of the 11-dimensional vielbein break up
as follows:
\be
e^A_M \rightarrow \left( \matrix{e^A_M&A_M\cr
                                  0&\phi\cr }
\right)
\quad ,
\ee
where we have set the $1\times 10$ block on the
lower left to zero using the freedom of those local
Lorentz transformations which mix the $x^{11}$ direction
with the other ten. The 10-dimensional massless vector
$A_M$ arises from the $e_M^{11}$ components of the vielbein,
while the real scalar $\phi$ (the 10-dimensional dilaton)
arises from the $e^{11}_{11}$ component.

The 11-dimensional antisymmetric tensor field $A_{MNP}$
splits into a 10-dimensional 3rd rank antisymmetric tensor
$A_{MNP}$ and a 2nd rank antisymmetric tensor $B_{MN}$
(from $A_{MN11}$). The 11-dimensional gravitino field
$\psi ^a_M$ becomes:
\be
\psi ^a_M \rightarrow \psi _M^{\alpha},\; \tilde{\psi}_M^{\alphadot},\;
\lambda ^{\alpha},\; \tilde{\lambda}^{\alphadot}\quad ;
\ee
giving two 10-dimensional Majorana-Weyl vector-spinors of
opposite chirality, and two 10-dimensional Majorana-Weyl
spinors of opposite chirality.

Since Type IIA supergravity is vectorlike it is trivially
free of gravitational anomalies. It is the field theory limit
of the Type IIA superstring.

\subsection{More on supersymmetry in 10 dimensions}

Type IIA supergravity has N=(1,1)$_{32}$ supersymmetry.
Since 16 is the minimum spinor dimension in ten dimensions,
we can in principle construct a chiral N=(1,0)$_{16}$
supergravity theory as well; however such a theory has
gravitational anomalies. The only other possibility in
ten dimensions is a chiral N=(2,0)$_{32}$ supergravity.
This theory, known as {\bf Type IIB supergravity},
turns out to be anomaly-free, due to highly nontrivial
cancellations. Type IIB cannot be obtained by dimensional
reduction of 11-dimensional supergravity.

N=(2,0)$_{32}$ supersymmetry is generated by two
Majorana-Weyl spinors $Q_{\alpha}$, $\tilde{Q}_{\alpha}$
of the same chirality. The on-shell Type IIB supergravity
multiplet has the following field content:
\begin{itemize}
\item $e^A_M$, the 10-dimensional vielbein.
\item $A_{MNPQ}$, a {\bf self-dual} 4th rank antisymmetric tensor.
\item $A_{MN}$, $\tilde{A}_{MN}$, 2nd rank antisymmetric tensors.
\item $A$, $\tilde{A}$, real scalars.
\item $\psi ^{\alpha}_M$, $\tilde{\psi}^{\alpha}_M$, Majorana-Weyl
vector-spinors of the same chirality.
\item $\lambda ^{\alphadot}$, $\tilde{\lambda}^{\alphadot}$,
Majorana-Weyl spinors of the same chirality.
\end{itemize}
Because the field content includes the self-dual part
of a 4th rank antisymmetric tensor, Type IIB supergravity
does not have a Lorentz covariant Lagrangian formulation.

The little group in 10-dimensions is $SO(8)$. Because of the
special automorphism symmetry of the Lie algebra $D_4$,
all of the dimension$<224$ irreps except the adjoint
occur as triplets of irreps with the same dimension and index.
These irreps are distinguished by subscripts $v$, $s$, and $c$.
Thus we can gain more information about the differences between
Type IIA and Type IIB supergravity by listing the $SO(8)$ irreps
corresponding to each component field. Note that both theories
have $128$$+$$128 =256$ real field components.

\begin{table}[t]
\caption{Comparison of Type II supergravities.
\label{tab:typetwo}}
\vspace{0.4cm}
\begin{center}
\begin{tabular}{|c|c|c|c|c|c|c|}
\hline
Type IIA:&1&28&$35_v$&$8_v$&$56_v$& \\
 &$\phi$&$B_{MN}$&$e^A_M$&$A_M$&$A_{MNP}$& \\
\cline{2-7}
 &$8_c$&$8_s$&$56_s$&$56_c$& & \\
 &$\lambda$&$\tilde{\lambda}$&$\psi$&$\tilde{\psi}$& & \\
\hline
Type IIB:&1&28&$35_v$&$1$&$28$&$35_c$\\
 &$A$&$A_{MN}$&$e^A_M$&$\tilde{A}$&$\tilde{A}_{MN}$&$A_{MNPQ}$\\
\cline{2-7}
 &$8_s$&$8_s$&$56_s$&$56_s$& & \\
 &$\lambda$&$\tilde{\lambda}$&$\psi$&$\tilde{\psi}$& & \\
\hline
\end{tabular}
\end{center}
\end{table}

Let us consider again 10-dimensional
N=(1,0)$_{16}$ supergravity, also known as
{\bf Type I supergravity}.
As we have noted, this theory is anomalous. Remarkably though, by coupling
this theory to the 10-dimensional supersymmetric Yang-Mills
multiplet, we can in certain cases obtain theories free
of both gravitational and mixed gauge-gravitational anomalies.
The field content of the chiral supergravity is just
a truncation of the Type IIA fields:
\begin{eqnarray*}
&&e^A_M,\;B_{MN},\;\phi \quad ;\\
&&\psi _M^{\alpha},\; \tilde{\lambda}^{\alphadot} \quad ;
\end{eqnarray*}
for a total of $64$$+$$64 =128$ real field components.
The 10-dimensional on-shell N=(1,0)$_{16}$ supersymmetric Yang-Mills
multiplet consists of a massless vector and a massless
Majorana-Weyl spinor
\[
A_M,\; \chi ^{\alpha}\quad ,
\]
in the $8_v$ and $8_c$ irreps of the little group, and the
adjoint representation of some gauge group.

The details \cite{chap} of how to couple N=(1,0) supergravity
to N=(1,0) Yang-Mills, consistent with both local
supersymmetry and gauge invariance, is beyond the
scope of these lectures. We merely note that one
surprising result is that the gauge invariant field
strength of the massless antisymmetric tensor field
$B_{MN}$ has an extra contribution which is a functional
of the Yang-Mills gauge field $A_M$. Then, even though
$B_{MN}$ is a gauge singlet, gauge invariance requires
that $B_{MN}$ transforms nontrivially under Yang-Mills
gauge transformations. This strange fact makes possible
the Green-Schwarz anomaly cancellation mechanism.

The coupled $d=10$ supergravity-Yang-Mills theory is
anomaly-free only for the following choices of gauge group:
\[
SO(32),\; E_8\times E_8,\; E_8\times [U(1)]^{248},\;
[U(1)]^{496}\quad .
\]
The first two choices correspond to the field theory limits
of the $SO(32)$ and $E_8\times E_8$ heterotic strings;
the other two choices can probably also be connected to
superstring theory using D-brane arguments.

\section{Conclusion}

It is a remarkable fact that many technical SUSY topics, thought
until recently to be of purely academic interest, have turned
out to be crucial to obtaining deep insights about the
physics of strongly-coupled nonabelian gauge theories and
strongly-coupled string theory. Even if there is no
weak scale SUSY in the real world -- even if there
is no SUSY at all -- supersymmetry has earned its place
in the pantheon of Really Good Ideas. Furthermore, we are
encouraged to continue to develop and expand the technical
frontiers of supersymmetry, confident both that there is still
much to learn, and that this new knowledge will find application
to important physical problems.


\section*{Acknowledgments}
The author is grateful to the TASI organizers for
scheduling his lectures at 9 a.m, to Jeff Harvey for
keeping me apprised of the Bull's playoff schedule, and to Persis
Drell for explaining to me what a minimum bias event is.
Fermilab is operated by the Universities Research Association, Inc.,
under contract DE-AC02-76CH03000 with the U.S. Department of Energy.

\section*{Appendix}
\subsection*{Notation and conventions}

My notation and conventions in these lectures
conforms with Wess and Bagger~\cite{wb} {\bf with the following
exceptions:}
\begin{itemize}
\item I use the standard ``West Coast'' metric:
\be
\metric = \left(\matrix{1&&&\cr
	 		&-1&&\cr
			&&-1&\cr
			&&&-1\cr}\right)
\label{eq:westcoast}
\ee
This is the standard metric convention of particle physics.
Wess and Bagger use the East Coast metric still popular with
relativists and other benighted souls.
\vskip .2in
Changes which follow from my different choice of metric:
\begin{enumerate}
\item My $\sigma ^0$ and $\barsigma ^0$ are the 2x2 identity matrix
instead of minus it.
\item I define the SL(2,C) generators with an extra factor of $i$:
\be
{\sigma ^{mn}_{\alpha}}^{\beta} = {i\over 4} \left[
\sigma ^{m}_{\alpha\gammadot}\barsigma ^{n\gammadot\beta}
- \sigma ^{n}_{\alpha\gammadot}\barsigma ^{m\gammadot\beta}\right]
\qquad .
\label{eq:slgens}
\ee
\item I define the Levi-Civita tensor density $\epsilon _{0123}
= +1$ instead of minus one.
\item As a result of this, $\sigma _{mn}$ is self-dual and
$\barsigma _{mn}$ is anti-self-dual, instead of vice-versa.
\item In the full component expansion of the chiral superfield
$\Phi(x,\theta,\bartheta)$, the
$(\theta\theta)(\bartheta\bartheta)$ term comes in with the
opposite sign.
\end{enumerate}
\end{itemize}
\subsection*{Spinor definitions and identities}
\bigskip
Irreps of SL(2,C) $\approx$ SO(1,3):
\bea
(\half,0) &= &{\rm left{\rm -}handed\ 2\ component\ Weyl\ spinor}\nonumber\\
(0,\half ) &= &{\rm right{\rm -}handed\ 2\ component\ Weyl\ spinor}\nonumber
\eea
In Van der Waerden notation, undotted = $(\half,0)$, dotted
= $(0,\half)$:
\bea
(\half,0)&\quad : \quad &\psid \qquad ,\nonumber\\
(0,\half )&\quad : \quad &\psiudot \equiv (\psid )^*
\qquad .\nonumber
\eea
Also:
\be
\psiddot \equiv (\psid )^{\dagger}\quad ;
\qquad \psiu \equiv (\psiddot )^* \qquad .
\label{eq:spinorconj}
\ee
\vskip .2in \ni
We raise and lower spinor indices with the 2-dimensional
Levi-Civita symbols:
\bea
\epsd = &\epsddot = &\left(\matrix{0&-1\cr
		                 1&0\cr}\right)
\quad ;\nonumber\\
\epsu = &\epsudot = &\left(\matrix{0&1\cr
		                              -1&0\cr}\right)
= i\sigma ^2 \quad .
\label{eq:levidef}
\eea
\bigskip
Thus:
\bea
\psiu &= \epsu \psi _{\beta} \quad ;\qquad \psid &= \epsd \psi ^{\beta}
\quad ;\nonumber\\
\psiudot &= \epsudot \barpsi _{\betadot} \quad ;
\qquad \psiddot &= \epsddot \barpsi ^{\betadot} \quad .
\label{eq:raiselower}
\eea
\medskip
Note also:
\bea
\epsd \epsilon ^{\beta\gamma} &= &\delta _{\alpha}^{\gamma}
\quad ,\nonumber\\
\epsddot \epsilon ^{\betadot\gammadot}
&= &\delta _{\alphadot}^{\gammadot} \quad ,\\
\epsd \epsilon ^{\delta\gamma} &=
&\delta _{\alpha}^{\gamma}\delta _{\beta}^{\delta} -
\delta _{\alpha}^{\delta}\delta _{\beta}^{\gamma}
\quad ,\nonumber\\
\epsddot \epsilon ^{\deltadot\gammadot} &=
&\delta _{\alphadot}^{\gammadot}\delta _{\betadot}^{\deltadot} -
\delta _{\alphadot}^{\deltadot}\delta _{\betadot}^{\gammadot}
\quad .\nonumber
\label{eq:epsilondefs}
\eea
Pauli Matrices:
\medskip
\be
\sigma ^1 = \left(\matrix{0&1\cr 1&0\cr}\right) \qquad
\sigma ^2 = \left(\matrix{0&-i\cr i&0\cr}\right) \qquad
\sigma ^3 = \left(\matrix{1&0\cr 0&-1\cr}\right)
\ee
\vskip .2in \ni
{}From which we define:
\bea
\sigma ^{m} &= (I,\vec{\sigma}) &= \barsigma _{m}
\quad ,\nonumber\\
\barsigma ^{m} &= (I,-\vec{\sigma}) &= \sigma _{m} \quad ,
\label{eq:covsigmadef}
\eea
where $I$ denotes the 2x2 identity matrix. Note that
in these definitions ``bar'' {\bf does not} indicate complex
conjugation.
\vskip .2in \ni
\begin{center} \openup 2\jot
$\sigma ^{m}\;$ has undotted-dotted indices: $\;\sig$ \\
$\barsigma ^{m}\;$ has dotted-undotted indices: $\;\sigdot$ \\
\end{center}
\vskip .2in \ni
We also have the completeness relations:
\bea
\tr \sigma ^{m} \barsigma ^{n} &= &2\metric \quad ,\nonumber\\
\sig \barsigma _{m}^{\gammadot\delta} &=
&2\delta _{\alpha}^{\delta}\delta _{\betadot}^{\gammadot}
\quad .
\label{eq:completeness}
\eea
\bigskip
$\sigma ^{m}$ and $\barsigma ^{m}$ are related by the Levi-Civita
symbols:
\be
\sigdot = \epsilon ^{\alphadot\gammadot}\epsilon ^{\beta\delta}
\sigma ^{m}_{\delta\gammadot} \quad; \quad
\sig = \epsilon _{\deltadot\betadot}\epsilon _{\gamma\alpha}
\barsigma ^{m\deltadot\gamma} \quad .
\label{eq:raisecovsigma}
\ee
\vskip .2in \ni
It is occasionally convenient to do a ``fake" conversion
of an undotted to a dotted index or vice versa using the fact
that $\sigma ^{0}$ and $\barsigma ^{0}$ are just the identity
matrix:
\be
\psiu = (\barpsi _{\betadot})^*\barsigma ^{0\betadot\alpha}
\quad ;\qquad
\psiudot = (\psi _{\beta})^*\sigma ^{0\beta\alphadot} \quad .
\label{eq:fakedef}
\ee
\vskip .2in \ni
Because the Pauli matrices anticommute, i.e.
\be
\{ \sigma ^i,\sigma ^j\} = \delta ^{ij}I\qquad i,j=1,2,3
\label{eq:paulianti}
\ee
we have the relations:
\bea
\left( \sigma ^{m}\barsigma ^{n} +
\sigma ^{n}\barsigma ^{m} \right) ^{\beta}_{\alpha} &=
&2\metric \delta _{\alpha}^{\beta} \quad ,\nonumber\\
\left( \barsigma ^{m}\sigma ^{n} +
\barsigma ^{n}\sigma ^{m} \right) ^{\betadot}_{\alphadot} &=
&2\metric \delta _{\alphadot}^{\betadot} \quad .
\label{eq:covsigmarel}
\eea
\vskip .2in \ni
{\bf Spinor Summation Convention:}
\bigskip
\bea
\psi\chi &= \psiu\chi_{\alpha} &= -\psid\chi^{\alpha} =
\chi^{\alpha}\psid = \chi\psi \nonumber\\
\barpsi\barchi &= \psiddot\barchi^{\alphadot}
&= -\psiudot\barchi_{\alphadot} =
\barchi_{\alphadot}\psiudot = \barchi\barpsi
\label{eq:spinorsumconv}
\eea
\vskip .1in
\noindent Note that these quantities are Lorentz scalars.

\noindent We also have:
\bea \openup 2\jot
(\chi\psi )^{\dagger} &= (\chi ^{\alpha}\psid )^{\dagger} =
\psiddot\barchi^{\alphadot} = \barchi\barpsi \nonumber\\
(\chi\sigma ^{m}\barpsi )^{\dagger} &=
\psi\sigma ^{m}\barchi = {\rm Lorentz\ vector}
\label{eq:spinorsumrels}
\eea
\vskip .2in
Other useful relations are:
\bea
\psiu \psi ^{\beta} &= &-\half\epsu \psi\psi \quad ,\nonumber\\
\psid \psi _{\beta} &= &\half\epsd \psi\psi \quad ,\\
\psiudot \barpsi ^{\betadot} &= &\half\epsudot \barpsi\barpsi
\quad ,\nonumber\\
\psiddot \barpsi _{\betadot} &= &-\half\epsddot \barpsi\barpsi
\quad .\nonumber
\label{eq:otheruseful}
\eea
\bigskip
The SL(2,C) generators are defined as
\bea
{\sigma ^{mn}_{\alpha}}^{\beta} &= &{i\over 4} \left[
\sigma ^{m}_{\alpha\gammadot}\barsigma ^{n\gammadot\beta}
- \sigma ^{n}_{\alpha\gammadot}\barsigma ^{m\gammadot\beta}\right]
\nonumber\\
{\hbox{$\barsigma ^{mn\alphadot}$}}_{\betadot} &= &{i\over 4} \left[
\barsigma ^{m\alphadot\gamma}\sigma ^{n}_{\gamma\betadot}
- \barsigma ^{n\alphadot\gamma}\sigma ^{m}_{\gamma\betadot}\right]
\label{eq:sltwocdef}
\eea
\vskip .2in
\noindent We have:
\bea
\epsilon ^{mnpq} \sigma _{pq} &= 2i\sigma ^{mn}\qquad ;
\quad&{\rm self\ dual\ } (1,0) \nonumber\\
\epsilon ^{mnpq} \barsigma _{pq} &= -2i\barsigma ^{mn}\qquad ;
\quad&{\rm anti\ self\ dual\ } (0,1)
\label{eq:selfdualdef}
\eea
\vskip .1in
\noindent And thus the trace relation:
\be
\tr \left[ \sigma ^{mn}\sigma ^{pq} \right] =
\half \left( \eta ^{mp}\eta ^{nq} -
\eta ^{mq}\eta ^{np} \right) + {i\over 2}
\epsilon ^{mnpq} \quad .
\label{eq:tracereldef}
\ee
\vskip .4in
\subsection*{Fierz identities}
\bigskip
\bea
(\theta\phi)(\theta\psi) &= &-\half (\theta\theta)
(\phi\psi)\\
& & \nonumber\\
(\bartheta\barphi)(\bartheta\barpsi) &= &-\half
(\barphi\barpsi)(\bartheta\bartheta)\\
& & \nonumber\\
\phi\sigma ^{m}\barchi &= &-\barchi\barsigma ^{m}\phi \\
& & \nonumber\\
\phi\sigma _{m}\barchi &= &-\barchi\barsigma _{m}\phi \\
& & \nonumber\\
(\theta\sigma ^{m}\bartheta)(\theta\sigma ^{n}\bartheta)
&= &\half \metric (\theta\theta)(\bartheta\bartheta)
\label{eq:goodfierz}\\
& & \nonumber\\
(\sigma ^{m}\bartheta)_{\alpha}(\theta\sigma ^{n}
\bartheta) &= &\half \metric \theta _{\alpha}(\bartheta\bartheta)
-i(\sigma ^{mn}\theta )_{\alpha}(\bartheta\bartheta)
\label{eq:myfierz}\\
& & \nonumber\\
(\theta\phi)(\bartheta\barpsi) &= &\half
(\theta\sigma ^{m}\bartheta)(\phi\sigma _{m}\barpsi)\\
& & \nonumber\\
(\bartheta\barpsi)(\theta\phi) &= &\half
(\theta\sigma ^{m}\bartheta)(\phi\sigma _{m}\barpsi) =
(\theta\phi)(\bartheta\barpsi)
\label{eq:fierztable}
\eea
\newpage
\subsection*{General 4-dimensional SUSY algebra}
\bigskip
\bea\openup 2\jot
&\{ Q^A_{\alpha},\barQ _{\betadot B}\} \qquad &=\qquad
2\sigma ^{m}_{\alpha\betadot}P_{m}\delta ^A_B \\
& & \nonumber\\
&\{ Q^A_{\alpha},Q^B_{\beta}\} \qquad &=\qquad
\epsd a^{\ell AB}B_{\ell} \\
& & \nonumber\\
&\{ \barQ _{\alphadot A},\barQ _{\betadot B}\} \qquad &=\qquad
-\epsddot a^*_{\ell AB}B^{\ell} \\
& & \nonumber\\
&[Q^A_{\alpha},P_m] \quad &=\quad [\barQ ^A_{\alphadot},P_m]
\quad = \quad 0 \\
& & \nonumber\\
&[Q^A_{\alpha},M_{mn}] \qquad &=\qquad
{\hbox{$\sigma _{mn\alpha}$}}^{\beta}Q^A_{\beta} \\
& & \nonumber\\
&[\barQ ^{\alphadot}_A, M_{mn}] \qquad &=\qquad
{\hbox{$\barsigma _{mn}^{\alphadot}$}}_{\betadot}
\barQ ^{\betadot}_A \\
& & \nonumber\\
&[P_m, P_n] \quad &= \quad 0 \\
& & \nonumber\\
&[M_{mn}, P_p] \quad &=\quad i(\eta _{np}P_m - \eta _{mp}P_n)\\
& & \nonumber\\
&[M_{mn},M_{pq}] &= -i\big( \eta _{mp}M_{nq} - \eta _{mq}M_{np}
\nonumber\\ &&\qquad
-\eta _{np}M_{mq} + \eta_{nq}M_{mp}\big) \\
& & \nonumber\\
&[Q^A_{\alpha},B_{\ell}] \qquad &=\qquad
{\hbox{$S_{\ell}^A$}}_B Q^B_{\alpha} \\
& & \nonumber\\
&[\barQ _{\alphadot A}, B_{\ell}] \qquad &=\qquad
-\hbox{$S^*_{\ell A}$}^B \barQ _{\alphadot B} \\
& & \nonumber\\
&[B_{\ell}, B_k] \qquad &=\qquad i{\hbox{$C_{\ell k}$}}^j B_j \\
& & \nonumber\\
&[P_m, B_{\ell}] \quad &=\quad
[M_{mn}, B_{\ell}] \quad =\quad 0
\label{eq:gensusyalg}
\eea
Where the $a^{\ell}$ are antisymmetric matrices, and
$S_{\ell}$, $a_{\ell}$ must satisfy the intertwining
relation:
\be
{\hbox{$S_{\ell}^A$}}_C \; a^{CBk} \quad =\quad
-a^{ACk} \; {\hbox{$S^{*\ell}_C$}}^B
\label{eq:intertwinerel}
\ee
Note also the perverse but essential convention implicit
in Wess and Bagger:
\be
a^{\ell AB} \qquad =\qquad -a^{\ell}_{AB}
\label{eq:wbconv}
\ee
\subsection*{N=1 SUSY algebra in 4 dimensions}
\bigskip
\bea\openup 2\jot
&\{ Q_{\alpha},\barQ _{\betadot}\} \qquad &=\qquad
2\sigma ^{m}_{\alpha\betadot}P_{m} \\
&\{ Q_{\alpha},Q_{\beta}\} \quad &=\quad
\{ \barQ _{\alphadot},\barQ _{\betadot}\} \quad =\quad 0\\
&[Q_{\alpha},P_m] \quad &=\quad [\barQ _{\alphadot},P_m]
\quad = \quad 0 \\
&[Q_{\alpha},M_{mn}] \qquad &=\qquad
{\hbox{$\sigma _{mn\alpha}$}}^{\beta}Q_{\beta} \\
&[\barQ ^{\alphadot}, M_{mn}] \qquad &=\qquad
{\hbox{$\barsigma _{mn}^{\alphadot}$}}_{\betadot}
\barQ ^{\betadot} \\
&[P_m, P_n] \quad &= \quad 0 \\
&[M_{mn}, P_p] \quad &=\quad i(\eta _{np}P_m - \eta _{mp}P_n)\\
&[M_{mn},M_{pq}] &= -i\left( \eta _{mp}M_{nq} - \eta _{mq}M_{np}
-\eta _{np}M_{mq} + \eta_{nq}M_{mp}\right) \\
&[Q_{\alpha},R] \qquad &=\qquad
RQ_{\alpha} \\
&[\barQ _{\alphadot}, R] \qquad &=\qquad
-R\barQ _{\alphadot} \\
&[P_m, R] \quad &=\quad
[M_{mn}, R] \quad =\quad 0
\label{eq:nonesusyalg}
\eea


\section*{References}
\begin{thebibliography}{99}
\bibitem{report}J. Amundson et al, ``Report of the
Supersymmetry Theory Subgroup'', hep-ph/9609374;
S. Dawson, ``SUSY and Such'', hep-ph/9612229;
M. Drees, ``An Introduction to Supersymmetry'', hep-ph/9611409;
J. Bagger, ``Weak Scale Supersymmetry: Theory and Practice'',
Lectures at TASI 95, hep-ph/9604232; X. Tata, ``Supersymmetry:
Where it is and How to Find it'', Talk at TASI 95,
hep-ph/9510287; H. Baer et al, ``Low-energy Supersymmetry
Phenomenology'', hep-ph/9503479.
\bibitem{wb}\Book{Supersymmetry and Supergravity}{J. Wess and J. Bagger}
{2nd edition, Princeton University Press, Princeton NJ, 1992.}
\bibitem{west}\Book{Introduction to Supersymmetry and Supergravity}
{Peter West}{2nd edition, World Scientific, Singapore, 1990.}
\bibitem{soh}M. Sohnius, {\bf Introducing Supersymmetry},
\Journal{\PREP}{128}{39}{1985}.
\bibitem{wit}N. Seiberg and E. Witten,
\Journal{\NPB}{426}{19}{1994}.
\bibitem{coleman}S. Coleman and J. Mandula, \Journal{\PR}{159}{1251}{1967}.
\bibitem{hls}R. Haag, J. {\L}opusza\'nski, and M. Sohnius,
\Journal{\NPB}{88}{257}{1975}.
\bibitem{vandam}For an interesting attempt, see H. van Dam and
L. Biedenharn, \Journal{\PLB}{81}{313}{1979}.
\bibitem{zumino}B. Zumino, \Journal{\JMP}{3}{1055}{1962}.
\bibitem{bps}See Jeff Harvey's lectures in this volume.
\bibitem{fsz}S. Ferrara, C. Savoy, and B. Zumino,
\Journal{\PLB}{100}{393}{1981}.
\bibitem{ferrara}S. Ferrara and B. Zumino, \Journal{\NPB}{79}{413}{1974}.
\bibitem{rocek}M. Grisaru, M. Rocek, and W. Siegel,
\Journal{\NPB}{159}{429}{1979}.
\bibitem{wz}J. Wess and B. Zumino, \Journal{\NPB}{70}{39}{1974}.
\bibitem{friedan}D. Friedan, \Journal{\PRL}{45}{1057}{1980};
M. Rocek, \Journal{\PHYSICA}{15}{75}{1985}.
\bibitem{zumsig} B. Zumino, \Journal{\PLB}{87}{203}{1979}.
\bibitem{supertwo}R. Grimm, M. Sohnius, and J. Wess,
\Journal{\NPB}{133}{275}{1978}; A. Galperin, E. Ivanov,
S Kalitzin, V. Ogievetsky, and E. Sokatchev,
\Journal{\CQG}{1}{469}{1984}; N. Ohta, H. Sugata, and
H. Yamaguchi, \Journal{\ANN}{172}{26}{1986};
P. Howe, ``Twistors and Supersymmetry'', hep-th/9512066.
\bibitem{andy}A. Strominger, \Journal{\CMP}{133}{163}{1990}.
\bibitem{luis}L. Alvarez-Gaume and D. Freedman,
\Journal{\CMP}{80}{443}{1981}.
\bibitem{sugra}P. van Nieuwenhuizen, {\bf Supergravity},
\Journal{\PREP}{68}{189}{1981}.
\bibitem{oldsei}N. Seiberg, \Journal{\PLB}{206}{75}{1988}.
\bibitem{shif}M. Shifman and A. Vainshtein, \Journal{\NPB}{277}{456}{1986}.
\bibitem{alv}L. Alvarez-Gaume and E. Witten,
\Journal{\NPB}{234}{269}{1983}.
\bibitem{julia}E. Cremmer, B. Julia, and J. Scherk,
\Journal{\PLB}{76}{409}{1978}.
\bibitem{green}\Book{Superstring Theory}{M. Green, J. Schwarz, and
E. Witten}{Cambridge University Press, Cambridge, 1987.}
\bibitem{chap}G. Chapline and N. Manton, \Journal{\PLB}{120}{105}{1983}.

\end{thebibliography}
\end{document}